\documentclass[a4paper,12pt]{article}

\usepackage{jheppub} 
\usepackage[export]{adjustbox}
\usepackage[utf8]{inputenc}

\usepackage{slashed}
\usepackage{bbm}
\usepackage{bm}
\usepackage{subfig}

\usepackage[normalem]{ulem}

\usepackage{soul}

\usepackage{caption}

\usepackage{pifont}

\usepackage{tikz}
\usepackage{tkz-euclide}
\usetikzlibrary{shapes.misc}
\usetikzlibrary {arrows.meta}
\usetikzlibrary{decorations.pathmorphing}	
\tikzset{
    v/.style={decorate, decoration={snake, segment length=3mm, amplitude=0.75mm}, draw},
    f/.style={draw=black, postaction={decorate},
        decoration={markings,mark=at position .6 with {\arrow[very thick]{latex}}}},
    fb/.style={draw=black, postaction={decorate},
        decoration={markings,mark=at position .4 with {\arrowreversed[very thick]{latex}}}},
    fnar/.style={draw=black},
    g/.style={decorate, draw=black,
        decoration={coil,amplitude=3pt, segment length=3.5pt}},
    s/.style={dashed,draw=black, postaction={decorate},
        decoration={markings,mark=at position .55 with {\arrow[very thick]{latex}}}},
    sb/.style={dashed,draw=black, postaction={decorate},
        decoration={markings,mark=at position .55 with {\arrowreversed[draw=black,very thick]{latex}}}},
    snar/.style={dashed,draw=black,line width =1.25pt},
    cross/.style={cross out, draw=black, minimum size=2*(#1-\pgflinewidth), inner sep=0pt, outer sep=0pt},
cross/.default={3pt},
}
\usepackage{multirow}
\graphicspath{{Figures/}} 

\newcommand{\al}[1]{\begin{align}\begin{aligned} #1 \end{aligned}\end{align}}

\def\be{\begin{equation}}
\def\ee{\end{equation}}

\title{Gravitational tests of electroweak relaxation}

\author[a,b]{Daniele Barducci}
\author[c]{Enrico Bertuzzo}
\author[c]{Mart\'in Arteaga Tupia}

\affiliation[a]{Universit\`a degli Studi di Roma la Sapienza, Piazzale Aldo Moro 5, 00185, Roma, Italy}
\affiliation[b]{INFN Section of Roma 1, Piazzale Aldo Moro 5, 00185, Roma, Italy}
\affiliation[c]{Instituto de F\'{i}sica, Universidade de S\~{a}o Paulo, C.P. 66.318, 05315-970 S\~{a}o Paulo, Brazil}

\emailAdd{daniele.barducci@roma1.infn.it}
\emailAdd{bertuzzo@if.usp.br}
\emailAdd{martin77@if.usp.br}

\abstract{
We consider a scenario in which the electroweak scale is stabilized via the relaxion mechanism during inflation, focussing on the case in which the back-reaction potential is generated by the confinement of new strongly interacting vector-like fermions. If the reheating temperature is sufficiently high to cause the deconfinement of the new strong interactions, the back-reaction barrier then disappears and the Universe undergoes a second relaxation phase. This phase stops when the temperature drops sufficiently for the back-reaction to form again. 
We identify the regions of parameter space in which the second relaxation phase does not spoil the successful stabilization of the electroweak scale.
In addition, the generation of the back-reaction potential that ends the second relaxation phase can be associated to a strong first order phase transition.
We then study when such transition can generate a gravitational wave signal in the range of detectability of future interferometer experiments.
}

\begin{document} 

\hspace{15cm}\vspace{-0.8cm}{\includegraphics[width=.08\textwidth]{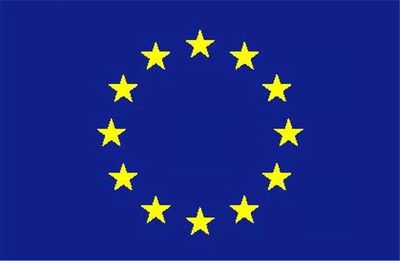}}

\maketitle

\section{Introduction}

The 2010s decade has been marked by two scientific milestones: the Higgs discovery at the Large Hadron Collider (LHC) in 2012 by the ATLAS and CMS collaborations~\cite{Aad:2012tfa,Chatrchyan:2012ufa} and the first direct detection of gravitational waves (GW) on Earth in 2015 by the LIGO and VIRGO collaborations~\cite{Abbott:2016blz, Abbott_2017}. While the latter has given access to previously unaccessible phenomena, like the merging of black holes binary systems, the first discovery has exacerbated the hierarchy problem, {\emph{i.e.}} the question of how the electroweak (EW) scale can be so much smaller than the Standard Model (SM) cutoff without the need for a large degree of fine tuning. Traditional symmetry based solutions like supersymmetry and composite dynamics are nowadays pushed in quite tuned regions of parameter space by the null LHC searches.

This has motivated the scientific community to consider
alternative solutions to the problem of the instability of the EW scale. A compelling possibility is the one where the Higgs mass is driven to a value much smaller than the SM cutoff by a dynamical evolution in the early Universe. This mechanism has been firstly  proposed in~\cite{Graham2015} and goes under the name of {\emph{cosmological relaxation}}. The basic idea is as follows: the Higgs squared mass parameter $H^\dag H$ is made dynamical by its coupling with a new scalar degree of freedom, the relaxion $\phi$, generally assumed to be a pseudo Nambu Goldstone boson (pNGB). The evolution of the relaxion field during the early Universe  evolution, governed by an opportune potential $V(\phi)$, scans the Higgs mass parameter, making it evolving from large positive values up to the critical value in which electroweak symmetry breaking (EWSB) is triggered. Once the Higgs develops a vacuum expectation value (VEV), a back-reaction potential turns on and stops the relaxion evolution, dynamically selecting the measured value for the EW scale.

 For the mechanism to work, two ingredients are essential: {\emph{i}}) a friction mechanism that slows down the relaxion evolution and avoids the overshooting of the back-reaction barrier and consequently of the correct EW scale, and {\emph{ii}}) a mechanism to generate the back-reaction itself.
In much of the explicit realizations of the relaxion mechanism, the friction is provided by the Hubble expansion during inflation~\cite{Graham2015,Espinosa:2015eda,Patil:2015oxa,Hardy:2015laa,Jaeckel:2015txa,Gupta:2015uea,Matsedonskyi:2015xta,Marzola:2015dia,DiChiara:2015euo,Ibanez:2015fcv,Fowlie:2016jlx,Kobayashi:2016bue,Choi:2016luu,Flacke:2016szy,Nelson:2017cfv,Jeong:2017gdy,Davidi:2017gir,Davidi:2018sii,Abel:2018fqg,Gupta:2019ueh}~\footnote{Notice that usually the details of the inflation sector are left largely unspecified. See however~\cite{Higaki:2016cqb,You:2017kah} for attempts to take into account inflaton effects, or~\cite{Tangarife:2017vnd,Tangarife:2017rgl} for an example of how to identify the relaxion with the inflaton.}, so that the relaxion field slow rolls during the cosmological relaxation phase. Alternatives are however possible: the friction can be generated by particle production~\cite{Hook:2016mqo,You:2017kah,Son:2018avk,Fonseca:2018kqf,Kadota:2019wyz}, the relaxion can fast-roll during inflation~\cite{Ibe:2019udh}, it can be stopped by a potential instability~\cite{Wang:2018ddr}, or it can fragment~\cite{Fonseca:2019ypl,Fonseca:2019lmc}. Cosmological relaxation after inflation is also possible~\cite{Fonseca:2018xzp}. As for the back-reaction, we can essentially distinguish between familon models~\cite{Gupta:2015uea} and models in which the potential barrier is generated by the confinement of a new strongly interacting dynamics with new vector fermions, as already discussed in the original paper~\cite{Graham2015}.
 
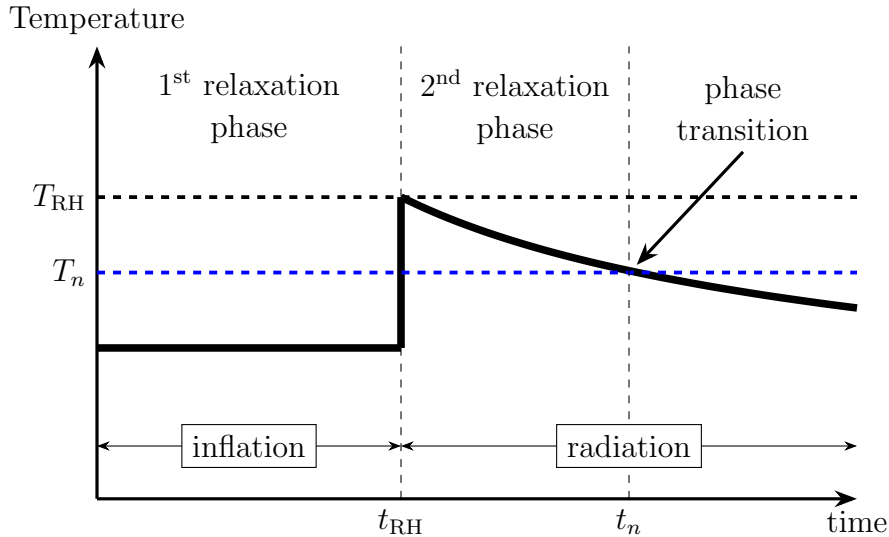
\begin{figure}
\begin{center}
\adjustbox{valign=m}{
 \begin{tikzpicture}[>=Stealth]
\draw[->,very thick] (0,0) -- (10,0) node[right,below]{time};
\draw[->, very thick] (0,0) -- (0, 6) node[above]{Temperature};
\draw[line width=1mm] (0,2) -- (4,2);
\draw[scale=1, smooth, domain=4:10, variable=\x, line width = 1mm] plot ({\x}, {8*\x^(-0.5)});
\draw[line width = 1mm] (4,2) -- (4,4);
\draw[dashed] (4,0) -- (4,6);
\draw[dashed, blue,line width = 0.5mm] (0, 3) -- (10, 3);
\draw[dashed] (7,0) -- (7,6);
\draw[<->] (0,0.7) -- (4,0.7);
\node[draw, fill=white] at (2,0.7){inflation};
\draw[<->] (4,0.7) -- (10, 0.7);
\node[draw, fill=white] at (7,0.7){radiation};
\node at (4,0) [below]{$t_{\rm RH}$};
\node at (7,0) [below]{$t_n$};
\node at (0,3) [left]{$T_n$};
\node[align=center] at (2, 5.2) {1$^{\rm st}$ relaxation \\phase};
\node[align=center] at (5.5, 5.2) {2$^{\rm nd}$ relaxation \\phase};
\draw[dashed, line width = 0.5mm] (0,4) -- (10, 4);
\node at (0, 4) [left]{$T_{\rm RH}$};
\draw[->,very thick] (8.5,4.6) node[above, align=center]{phase \\ transition}-- (7.1, 3.1);
\end{tikzpicture}} 
\end{center}
\caption{\label{fig:summary} Sketch of the framework we are considering. The thick black line represents an approximate sketch of the temperature evolution. During inflation, a first relaxation phase selects the correct EW vacuum. When reheating happens at $t_{{\rm RH}}$, the temperature increases and the radiation domination phase begins. For simplicity, we will consider reheating to be an instantaneous process. If the reheating temperature $T_{{\rm RH}}$ is larger than the nucleation temperature $T_n$, the back-reaction potential disappears and a second relaxation phase occurs. After a while, the temperature falls below the nucleation temperature of the new strongly interacting group, and a phase transition occurs which may generate a GW signal.}
\end{figure}

In this work, we focus on the second scenario. This opens up an interesting possibility: after a relaxation phase during inflation in which the EW scale is dynamically selected, the Universe may be reheated to temperatures above the critical temperature of the new confining interactions. If this happens, the back-reaction barrier disappears and the Universe undergoes a second relaxation phase. When the temperature of the Universe drops again below the confinement scale of the new strong dynamics, the barrier is once again generated and the relaxion stops again its evolution. Crucially, depending on the gauge group of the new confining dynamics, the number of new fermions and their representations under the gauge group, this phase transition can be of first order~\cite{Pisarski:1983ms} and can thus give rise to a stochastic GW background~\cite{Witten:1984rs}. The signal might be then detected at present and future interferometer experiments~\cite{Caprini_2016, Caprini:2019egz, Kuroda_2015, NI_2013, Ni_2016,Badurina_2020, graham2017midband,Maggiore_2020,reitze2019cosmic}, thus connecting in this way the two milestones discovery of the 2010s. A sketch of the situation we are considering is shown in Fig.~\ref{fig:summary}. 

The paper is organized as follows: in Section~\ref{sec:relaxation_inflation} we review the cosmological relaxation mechanism, where the back-reaction potential is generated by a new strongly interacting dynamics. In Section~\ref{sec:evolution_after_reheating} we analyze the relaxion evolution after reheating. We show how the equation of motion of the relaxion during this second relaxation phase can be analytically solved for a certain range of temperatures, and what are the bounds on the parameter space that we obtain by requiring the additional relaxation phase not to spoil the dynamical selection of the EW scale achieved during the first relaxation phase.
 In Section~\ref{sec:GW_signal} we discuss the GW spectrum generated in the first order phase transition associated with the generation of the back-reaction potential after reheating and its detectability at future experiments. 
  We thus conclude in Section~\ref{sec:concl}. We also add two Appendices where we collect more technical material for the interested reader. In Appendix~\ref{sec:strong_int_models} we discuss in detail various models of strongly interacting vector-like fermions and review how to describe their low energy dynamics 
 and in particular their vacuum energy, which is relevant for the relaxion mechanism. In  Appendix~\ref{sec:succ_relaxation} we instead explicitly show how the constraints on the relaxion parameter space used throughout the paper are derived.

\section{Relaxation with strongly interacting fermions}\label{sec:relaxation_inflation}

In this section we briefly summarize the relaxation mechanism, highlighting its main features. Following~\cite{Graham2015} we consider a scalar potential
\be\label{eq:potential}
V(H,\phi) = \left(\Lambda^2 - \epsilon \Lambda \phi \right) H^\dag H  + \lambda (H^\dag H)^2- r \epsilon \Lambda^3 \phi + V_{{\rm BR}}(H,\phi)\ ,
\ee
where $\Lambda$ is the cutoff of the theory, $\lambda$, $\epsilon$ and $r$ are positive dimensionless parameters and $V_{{\rm BR}}(H,\phi)$ is the back-reaction potential. The dependence of the first term 
of Eq.~\eqref{eq:potential} on $\phi$ makes the Higgs mass squared parameter a dynamical quantity.
As for the back-reaction potential, $V_{{\rm BR}}(H,\phi)$, we consider it to be of the form
\be
V_{{\rm BR}}(H,\phi) =  \Lambda_{{\rm{BR}}}^4(\langle H \rangle) \cos\left(\frac{\phi}{F}\right) \ ,
\ee
where $ \Lambda_{BR}^4(\langle H \rangle)$ is a (model dependent) monotonically increasing function of the Higgs VEV and $F$ is the scale at which the NGB $\phi$ appears. Notice that the back-reaction potential respects a discrete shift symmetry $\phi \to \phi + 2\pi F$. This symmetry is explicitly broken by the spurion $\epsilon$, which can thus be taken small in a technically natural way. 

We assume that the cosmological relaxation mechanism takes place during inflation, with $\phi$ evolving in a slow-roll regime. We furthermore take the initial value of the field $\phi$ to be sufficiently small to guarantee the condition $\Lambda^2 - \epsilon \Lambda \phi_{in}>0$ to be satisfied at the beginning of the relaxation phase, thus implying an unbroken EW symmetry.

As $\phi$ evolves, increasing its value due to the $- r \epsilon \Lambda^3 \phi$ term in the potential, the Higgs squared mass parameter becomes smaller and smaller, until it crosses zero and causes the breaking of the EW symmetry. Once $\langle H \rangle \neq 0$, the back-reaction potential is switched on. The amplitude of the oscillating term grows with the Higgs VEV, until it's large enough to stop the evolution of $\phi$, dynamically selecting the value of $\langle H \rangle$. This is the essence of the relaxion mechanism (see Fig~\ref{fig:relaxion_higgs}).

A possible origin for the back-reaction potential, already considered in the original work~\cite{Graham2015}, involves fermions charged under new strong interactions $\mathcal{G}_{{\rm  dark}}$ as well as under EW interactions.
Provided that the scale of inflation is smaller that the confinement scale of $\mathcal{G}_{{\rm  dark}}$, which we will denote as $g_\rho f$, and that the Higgs boson interacts with the new fermions, the back-reaction potential forms and, as described in App.~\ref{sec:strong_int_models}, $\Lambda_{{\rm BR}}$ takes the form
\be\label{eq:back-reaction}
\Lambda_{{\rm BR}}^4 \simeq \left| \mu_B^2 H^\dag H - \Lambda_0^4 \right|\ ,
\ee
where the constants $\mu_B^2$ and $\Lambda_0^4$ depend on the specific model considered. The form of Eq.~\eqref{eq:back-reaction} is however generic. 
We conclude this section by listing the conditions that are needed to guarantee the successful realization of the relaxion mechanism, see Appendix~\ref{sec:succ_relaxation} for more details.
\begin{description}
\item[Validity of the EFT:] We are assuming the field $\phi$ to be an angular degree of freedom, with a decoupled radial mode. Since the mass of the radial mode is $\mathcal{O}(F)$ we obtain the condition $F \gtrsim \Lambda$. We will also require $F \lesssim M_{{\rm Pl}}$, where $M_{{\rm Pl}} \simeq 10^{18}\;$GeV is the Planck mass.
\item[Conditions on inflation:] We ask the dynamics of inflation to be decoupled from the one of the relaxion. To achieve this we require the relaxion energy density $\rho_\phi \sim \Lambda^4$ to be smaller than the inflaton energy density, $\rho_{{\rm infl}} \sim H_I^2 M_{{\rm{Pl}}}^2$, where $H_I$ is the Hubble scale during inflation. This implies the lower bound
\be\label{eq:lower_b_HI}
H_I \gtrsim \frac{\Lambda^2}{M_{{\rm{Pl}}}}\ .
\ee
Since we are assuming that the back-reaction is generated by some new strong force, we need to guarantee that the barriers can form during inflation. This requires
\be
H_I \lesssim \Lambda_d\ ,
\ee
where $\Lambda_d$ denotes the scale of confinement of the new strong interactions. 

We also demand the classical relaxion evolution not to be spoiled by quantum fluctuations. To achieve this condition we require the classical spread of the relaxion field in one Hubble time, $\Delta \phi_{{\rm cl}} \sim \dot{\phi}/H_I \sim V'/H_I^2$, to be larger than the quantum spread $\Delta \phi_{{\rm quantum}} \sim H_I$, where $V$ indicates the relaxion potential and the derivative is with respect to the field $\phi$. This implies a second upper bound on the scale of inflation that reads
\be
H_I \lesssim \epsilon^{1/3} \Lambda \ .
\ee
\item[Conditions on the back-reaction:] The term $\Lambda_0^4$ is present in the back-reaction potential even before EWSB. To guarantee that it does not stop the relaxion evolution we require
\be\label{eq:cond1}
\Lambda^3 > \frac{\Lambda_0^4}{r \epsilon F} \ .
\ee
In addition, to ensure that after EWSB there will be a period of evolution in which the height of the barrier grows, we need
\be\label{eq:cond2}
v^2_{{\rm EW}} \mu_B^2 > \Lambda_0^4 \qquad \mathrm{and} \qquad \Lambda < \frac{\mu_B^2}{\epsilon F}\ .
\ee

\item[The EW scale is an output of the relaxation:] An approximate expression for the EW VEV in terms of the parameters of the model is
\be
v_{{\rm EW}}^2 \simeq \frac{\epsilon \Lambda^3 F + \Lambda_0^4}{\mu_B^2 - \epsilon \Lambda F} \ .
\ee
\end{description}
As shown in Appendix~\ref{sec:succ_relaxation}, the cutoff satisfy 
\be\label{eq:upper_bound_Lambda}
\Lambda \lesssim   6.7\times 10^6\,\mathrm{GeV}~ \left[ \frac{\mu_B}{10\, {\rm GeV}} \right]^{2/3} \left[ \frac{10^{-30}}{\epsilon}\right]^{1/3} \left[ \frac{10^{16}\, {\rm GeV}}{F} \right]^{1/3} \left[ 1- \frac{\Lambda_0^4}{\mu_B^2 v^2}\right]^{1/3}\ ,
\ee
favoring thus very small values of $\epsilon$. This result can be problematic in two aspects: {\emph i)} to completely solve the hierarchy problem an additional protection mechanism must be present for scales between $\Lambda$ and $M_{{\rm Pl}}$, and {\emph{ii)}} a successful relaxation requires the relaxion excursion to be at least $\Delta \phi \sim \Lambda/\epsilon$ in order for the Higgs mass parameter to change sign. Given the very small $\epsilon$ needed for the mechanism to work, the resulting excursion is transplanckian. The first issue can be solved assuming supersymmetry~\cite{Batell:2015fma,Evans:2016htp,Evans:2017bjs}, a composite dynamics~\cite{Batell:2017kho} or a warped dimension~\cite{Fonseca:2017crh} to be present above $\Lambda$. The second problem requires more model building effort, but can be solved in the context of clockwork models~\cite{Choi:2015fiu,Kaplan:2015fuy,Giudice:2016yja,Craig:2017cda}. In the following we assume that one of these mechanisms is present to stabilize the EW scale all the way up to the Planck scale, and focus only on the effective field theory defined by Eq.~\eqref{eq:potential}. 

Let us conclude this section with a comment on the compatibility between the well-known upper bound on the reheating temperature, $T_{\rm RH} \lesssim \sqrt{H_I M_{PL}}$, and the conditions on the Hubble parameter during inflation needed for the relaxation mechanism to work. If we take $H_I$ close to the lower bound of Eq.~\eqref{eq:lower_b_HI} we see that $T_{\rm RH} \lesssim \Lambda$, i.e. the phase following reheating can be described inside the validity of the EFT considered. Since the value of the Hubble constant during reheating has no effect on our subsequent considerations, we will take $H_I \sim \Lambda^2/M_{PL}$ from now on.

\section{Evolution after reheating}\label{sec:evolution_after_reheating}

As already mentioned, we are considering a situation in which the relaxation of the EW scale happens during inflation, {\emph{i.e.}} the correct EW VEV is selected at the end of this phase. This leaves open the question of \textit{what happens after reheating?} One possibility is for reheating to leave the Universe in a bath with temperature $T_{{\rm RH}} \lesssim \Lambda_d$. In this case there is no further dynamical evolution in the relaxion field direction, since $\phi$ remains stuck in its minimum. This is what is implicitly assumed in the original work~\cite{Graham2015}. If $T_{\rm RH} > T_{\rm EW}$, with $T_{\rm EW}$ the scale of EW phase transition, we expect thermal corrections in the Higgs field direction to recover the EW symmetry. As the Universe expands and cools down, EWSB is again triggered as usual~\footnote{Since the portal coupling between the Higgs and the relaxion sectors is small, we do not expect significant modifications to the EW phase transition with respect to what happens in the SM.}. 

A second possibility, on which we focus in the following, is for reheating to happen at $T_{\rm RH} > {\rm max}(T_d, T_n)$, where $T_d$ and $T_n$ are the temperatures associated with the deconfinement and confinement of the additional strong interaction ${\cal G}_{{\rm dark}}$. If this is the case, after reheating the back-reaction disappears and the Universe undergoes a second period of relaxation. This possibility has received less attention from the literature, see {\emph{e.g.}}~\cite{Hardy:2015laa,Higaki:2016cqb,Choi:2016kke,Banerjee:2018xmn}. As the Universe expands and cools down, it will reach a temperature $T \sim T_n$ at which
the new strong sector again confines, thus producing again the back-reaction barrier and ultimately stopping the relaxion evolution~\footnote{Note that
the oscillations of the relaxion around the minimum can make it a viable candidate for dark matter~\cite{Banerjee:2018xmn}.}. As we are going to see at the end of this Section, the relaxion evolution stops soon after the barrier is again formed. If the transition producing the back-reaction barrier is strongly first order, it will produce a GW signal that might be observable at interferometer experiments, as we will study in Sec.~\ref{sec:GW_signal}. This allows to open a new window on this type of solutions of the hierarchy problem.

Let us now discuss more in detail the relaxion evolution when $T_{\rm RH} > {\rm max}(T_d, T_n)$. We want to understand whether this additional relaxation phase can spoil the solution of the hierarchy problem. We achieve this by imposing additional conditions on the parameters in order to achieve the correct EW minimum today.  We will also make the following simplifying assumptions: (i) reheating is instantaneous and (ii) the energy density of the relaxion after reheating is subdominant with respect to the energy density in radiation. All our assumptions aim at avoiding additional model building related to the reheating or the deconfinement phases~\footnote{Notice that, on general grounds, achieving deconfinement may require additional model building. See Ref.~\cite{Croon:2019ugf} for a recent example in a different context. Also, the order of the deconfinement phase must be carefully studied, as it is not obvious that it is of the same order as the confinement phase transition.}. We also ignore a potential gravity wave signal generated during the deconfining phase, since the spectrum would depend on the details of the process. While interesting, we leave the study of these problems to future work, focussing only on what happens after reheating. With our assumption that the barrier disappears when the universe enters the radiation domination phase after inflation we need to consider three situations, depending on the relative hierarchy between $T_{\rm RH}$, $T_n$ and $T_{\rm EW}$:
\begin{itemize}
\item If $T_n < T_{\rm RH} < T_{\rm EW}$ the Universe is reheated to a phase in which the scalar potential has already a non-trivial minimum in the Higgs direction, while there are no barriers in the relaxion direction. Using the potential in Eq.~\eqref{eq:potential} in the equation of motion $\ddot{\theta} + 3 H(t) \dot{\phi} + \partial V/\partial\phi = 0$ we obtain 
\be\label{eq:EoM_after_reheating}
\ddot{\theta} + \frac{3}{2t}\dot{\theta}  = \frac{1}{F^2} \bigg( r \epsilon F \Lambda^3  + \epsilon F \Lambda v^2(\theta) \bigg) \ ,
\ee
where 
\be
v^2(\theta) = \frac{-\Lambda^2 + \epsilon F \Lambda\, \theta}{2 \lambda}
\ee
is the Higgs VEV in the absence of the back-reaction barrier. We have used $H=1/(2t)$, as appropriate for a radiation dominated Universe. The solution for temperatures $T_n < T < T_{\rm RH}$ can be written analytically in terms of modified Bessel functions $I_n$ and $K_n$ of first and second kind, respectively~\footnote{In Wolfram Mathematica these functions are called BesselI and BesselK, respectively.}. To write a compact expression we define a new time variable $\tau = \epsilon \Lambda t/\sqrt{2\lambda}$, the functions
\be
f(\tau) = \frac{I_{1/4}(\tau)}{\tau^{1/4}}\ , ~~~~ g(\tau)  = \frac{K_{1/4}(\tau)}{\tau^{1/4}}\ ,
\ee
and the constant term
\be
\xi = \frac{(2 r \lambda - 1) \Lambda^2}{\sqrt{2 \lambda} F^2}\ .
\ee
The solution can be written as
\al{\label{eq:sol_vsq_neq_0}
\theta(\tau) & = -\xi + \frac{g_0 \theta_0^\prime - g_0^\prime (\theta_0 + \xi)}{f_0^\prime g_0 - f_0 g_0^\prime} f(\tau) +  \frac{-f_0 \theta_0^\prime + f_0^\prime (\theta_0 + \xi)}{f_0^\prime g_0 - f_0 g_0^\prime} g(\tau)\ ,
}
where the primes denote the derivative with respect to $\tau$, and the subscript~$0$ indicates that the quantity must be computed at the initial time, {\emph{i.e.}} at reheating. Since we are assuming reheating to be an instantaneous process, we can identify $\theta_0$ with the value of the relaxion field at the end of the inflationary period and we can assume that the relaxion field does not acquire a relevant dynamics during reheating, leading to $\dot\theta_0 \simeq 0$. We also notice that for large values of $\tau$ we have the asymptotic behavior $I_n(\tau) \sim e^\tau/\sqrt{\tau}$ and $K_n(\tau) \sim e^{-\tau}/\sqrt{\tau}$. For $\tau \gg 1$ we thus obtain 
\be
\theta(\tau) \simeq -\xi + \left(\frac{\tau_0}{\tau} \right)^{3/4} \cosh(\tau-\tau_0) \left( \theta_0 + \xi\right) + \left(\frac{\tau_0}{\tau} \right)^{3/4} \sinh(\tau-\tau_0)\,  \theta_0^\prime\ .
\ee
The dependence on hyperbolic trigonometric functions implies that the relaxion field will evolve very quickly, if enough time is allowed to pass. We will comment later on the dynamics taking place for $T < T_n$;

\item If $T_n < T_{\rm EW} < T_{\rm RH} $ there are two phases in the evolution. The first one applies for $T_{\rm EW} < T < T_{\rm RH}$ while the second one for $T_n < T < T_{\rm EW}$. In the first phase the equation of motion to be solved is Eq~\eqref{eq:EoM_after_reheating} with $v^2(\theta) = 0$. The solution can again be written analytically:
\be\label{eq:solution_afterRH_above_EWSB}
\theta(t) = \theta(T_{\rm RH}) + \frac{r \epsilon \Lambda^3}{5F} (t^2 - 5 T_{\rm RH}^2) + \frac{4 r \epsilon \Lambda^3 T_{\rm RH}^{5/2}}{5 F \sqrt{t}}\ .
\ee
Notice that the evolution is a power law in this regime, and the relaxion field evolves much less than what happens when $v^2(\theta) \neq 0$. 
Using the relation between time and temperature in a radiation domination Universe
\be
H^2 = \frac{1}{4\, t^2} = \frac{1}{3 M_{{\rm Pl}}^2} \frac{\pi^2}{30} g_\star T^4 \ ,
\ee
where $g_\star$  denotes the number of relativistic degrees of freedom we can rewrite the solution as
\be\label{eq:solution_afterRH_above_EWSB_2}
\theta(T) = \theta(T_{\rm RH}) + \frac{ 9}{2 \pi^2 g_\star} \frac{r \epsilon \Lambda^3 M_{PL}^2}{F} \frac{4 T^5 - 5 T^4 T_{\rm RH} + T_{\rm RH}^5}{T^4 T_{\rm RH}^5}\ .
\ee
Once EWSB is triggered the equation of motion to be solved is again Eq.~\eqref{eq:EoM_after_reheating} with $v^2(\theta) \neq 0$, whose solution is given in Eq.~\eqref{eq:sol_vsq_neq_0} with $\tau_0 \equiv \tau_{EW}$;
\item The last case is the one in which $T_{\rm EW} < T_n < T_{\rm RH}$. For $T_n < T < T_{\rm RH}$ the solution is given by Eq.~\eqref{eq:solution_afterRH_above_EWSB}, while we will discuss in the next paragraph what happens for $T < T_n$.
\end{itemize}
\begin{figure}[tb]
\includegraphics[width=\textwidth]{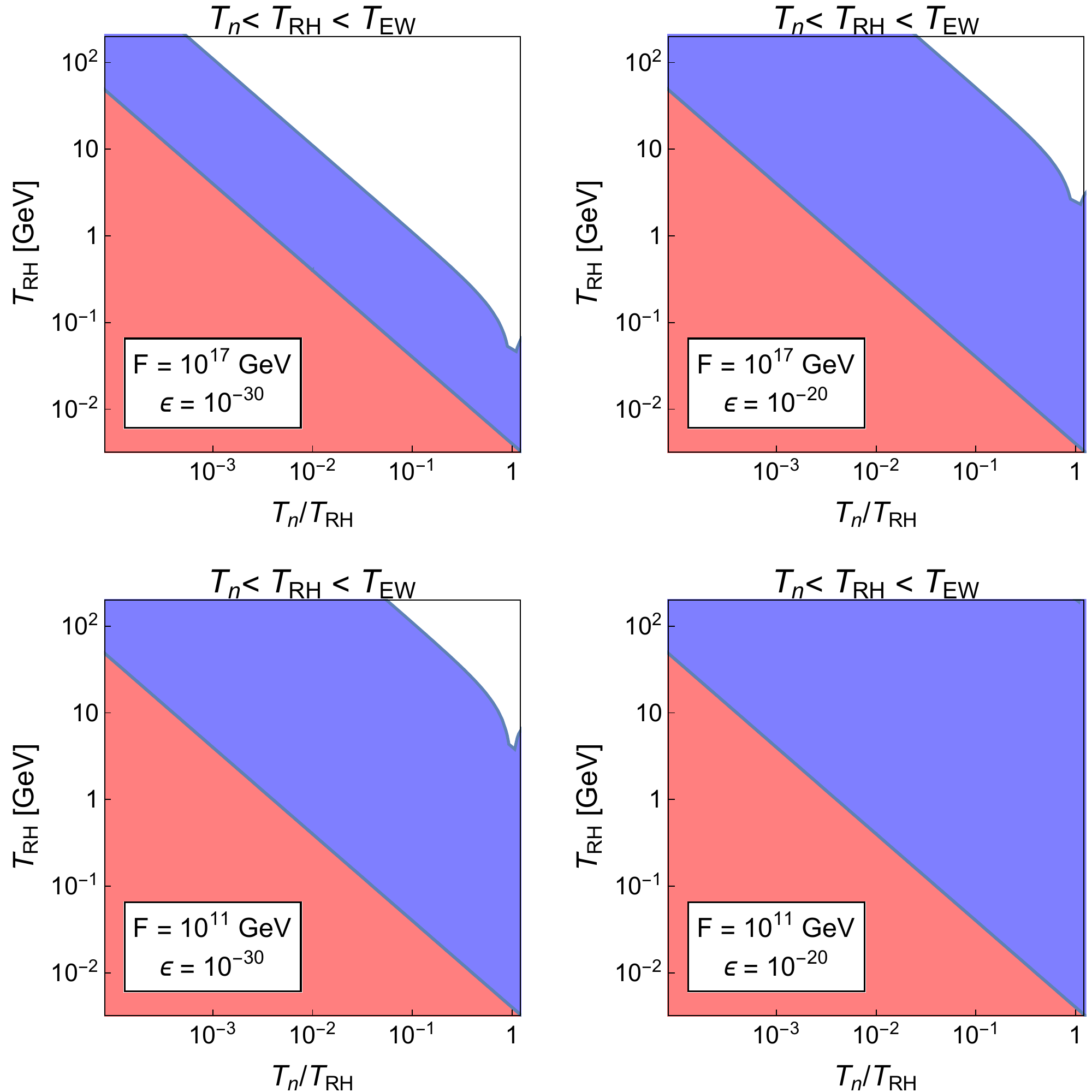}
\caption{\label{fig:vev_relative_difference_1} Regions in which the relative variation of the Higgs VEV of Eq.~\eqref{eq:v_relative_variation} is larger than unity (blue) and therefore the solution of the hierarchy problem is spoiled and in which the nucleation or the reheating temperatures are smaller than $4$ MeV (red). We suppose the initial VEV to be $v_{{\rm EW}}$. All the plots show the $T_n < T_{{\rm RH}}<T_{{\rm EW}}$ case. See the text for more details.}
\end{figure}
\begin{figure}[tb]
\includegraphics[width=\textwidth]{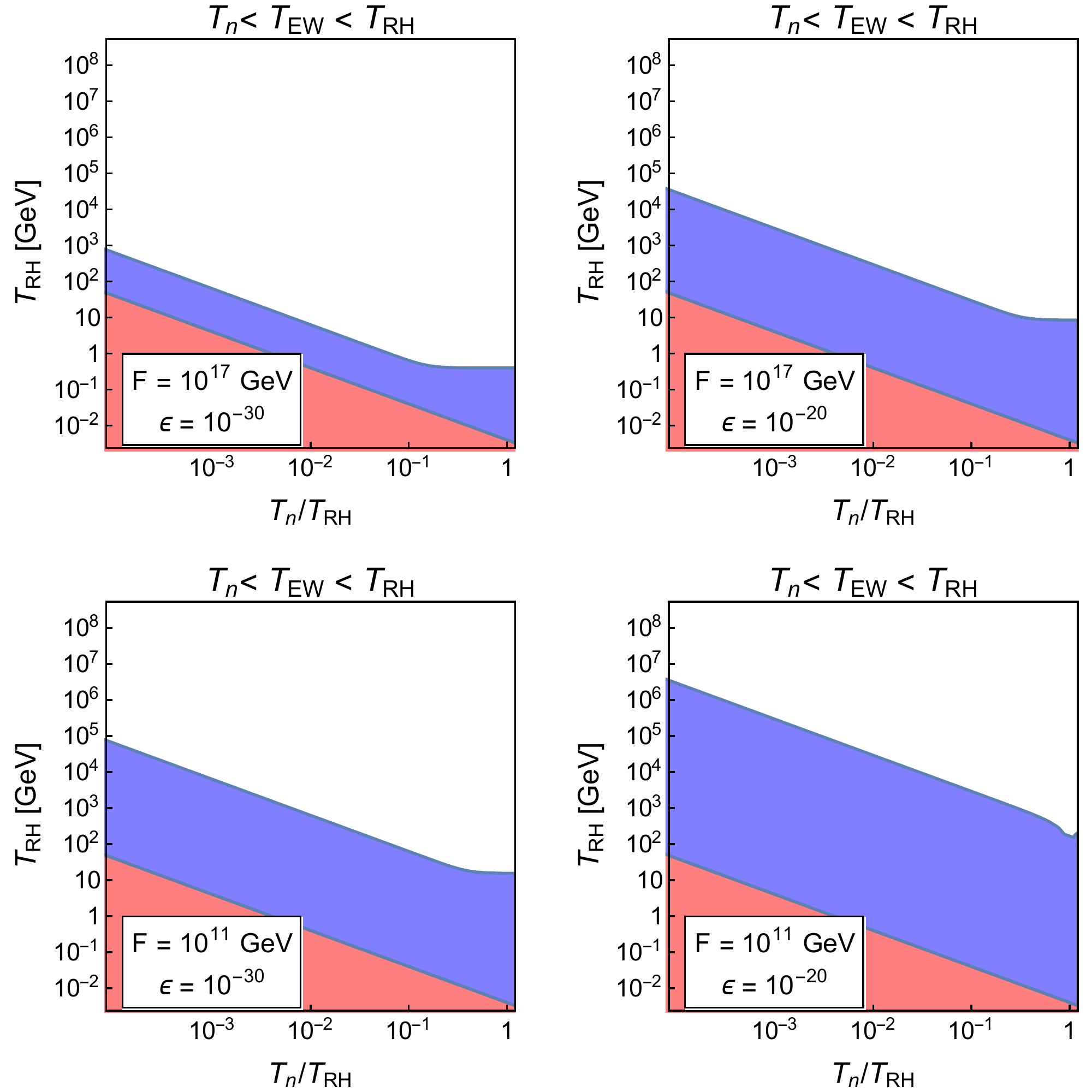}
\caption{\label{fig:vev_relative_difference_2}Regions in which the relative variation of the Higgs VEV of Eq.~\eqref{eq:v_relative_variation} is larger than unity (blue)  and therefore the solution of the hierarchy problem is spoiled and in which the nucleation or the reheating temperatures are smaller than $4$ MeV (red). We suppose the initial VEV to be $v_{{\rm EW}}$. All the plots show the  $T_n < T_{{\rm EW}}<T_{{\rm RH}}$  case. See the text for more details. }
\end{figure}
To compute the solution for $T < T_n$ we need to turn on not only the Higgs VEV, for $T<T_{\rm EW}$, but the back-reaction barrier as well. To the best of our knowledge, no analytical solution can be found in this case. We can however qualitatively expect that, once the barriers form, the relaxion will find itself trapped in a position that is displaced from the minimum in the relaxion direction. It will then oscillate around this minimum losing energy. It is in this phase that it can behave like dark matter, as studied in~\cite{Banerjee:2018xmn}. Whether or not the relaxion stops its evolution when encountering the first barrier
will depend on the velocity it has acquired during the second relaxation phase, and cannot be inferred analytically. As shown at the end of App.~\ref{sec:succ_relaxation} however, the change in the Higgs VEV between two subsequent minima is relatively small, so that it is likely that the second stopping phase will not modify dramatically the value of the Higgs VEV at the end of the relaxion evolution. What can modify dramatically this value is however the relaxion evolution before the barrier is again formed. To understand whether this is the case, we show in Fig.~\ref{fig:vev_relative_difference_1} and Fig.~\ref{fig:vev_relative_difference_2} in blue the regions in which
\be\label{eq:v_relative_variation}
\frac{\delta v}{v} \equiv \frac{v[\theta(T_n)] - v_{EW}}{v_{EW}} > 1 \ .
\ee
In these regions the Higgs VEV at the end of the second relaxation phase differs by more than a factor of $2$ with respect to the observed VEV, and the solution of the hierarchy problem is spoiled. 
In Fig.~\ref{fig:vev_relative_difference_1} we consider the case in which $T_n < T_{{\rm RH}} < T_{\rm EW}$, while in Fig.~\ref{fig:vev_relative_difference_2}  we show the case in which $T_n < T_{{\rm EW}} < T_{{\rm RH}}$. We take $T_{{\rm EW}} \simeq 160$ GeV. We have fixed the parameters of the model to the same representative values chose in Eq.~\eqref{eq:upper_bound_Lambda}, with the exceptions of $\epsilon$ and $F$, whose values are reported in the plots. For simplicity we have taken a fixed value of $g_\star = 100$, although by varying it the overall picture does not change. Furthermore, we have supposed that the initial relaxion value at reheating gives $v_{{\rm EW}}$ and has vanishing velocity. The quantity $v[\theta(T_n)]$ is computed using Eqs.~\eqref{eq:sol_vsq_neq_0} and~\eqref{eq:solution_afterRH_above_EWSB_2}. We also impose a lower bound $T_n, T_{{\rm RH}} \gtrsim 4\;$MeV, represented by the red regions, where the limit on $T_{\rm RH}$ is taken from~\cite{Hannestad:2004px} while the limit on $T_n$ is imposed to be safely far from the Big Bang nucleosynthesis epoch $T\sim  1\;$MeV.
By inspecting Fig.~\ref{fig:vev_relative_difference_1} we see that for reheating temperatures below the EW phase transition one, there are regions in parameter space in which the solution of the hierarchy problem is completely spoiled, depending on the choice of parameters. This is due to the exponential evolution of the relaxion field in this regime. On the contrary, when $T_n < T_{\rm EW} < T_{\rm RH}$ most of the relaxion evolution happens when Eq.~\eqref{eq:solution_afterRH_above_EWSB_2} is valid. Since the relaxion field does not evolve much in this regime, there are large regions of parameter space in which the modifications of the VEV are small. Finally, when $T_{{\rm EW}} < T_n < T_{RH}$ the evolution follows again Eq.~~\eqref{eq:solution_afterRH_above_EWSB_2} and is not enough to drastically modify the value of the VEV. For this reason we do not show any plot for this case. 

Let us finish this Section noticing that we can imagine a situation in which the VEV at the end of the first relaxation phase is smaller than the observed Higgs VEV. If this is the case, the second relaxation phase could provide the additional evolution needed to be compatible with experiments. In this paper we will focus on a situation in which the correct Higgs VEV is selected at the end of the first relaxation phase, leaving for future work the analysis of what happens in the opposite situation.

\section{Gravitational waves signal}\label{sec:GW_signal}

In relaxion models with strongly interacting fermions there are two main sources of GW: {\emph{i)}} the confinement of $\mathcal{G}_{{\rm dark}}$ if it proceeds through a first order phase transition~\cite{Helmboldt:2019pan, Croon:2019iuh} and {\emph{ii)}} the possible penetration of the barrier by the relaxion field before stopping at the minimum~\cite{Hebecker_2016,Brown_2016,zhou2020probing}.~\footnote{As noticed above, we ignore a possible signal coming from the deconfining of the strongly interacting group. We have estimated that, in ample regions of the parameter space shown in Figs.~\ref{fig:vev_relative_difference_1} and~\ref{fig:vev_relative_difference_2}, the time passed between deconfinement and confinement is sufficiently long to avoid interference between the GW signals generated in the two processes.} The last process is much more difficult to analyze we will focus here on the first signal, leaving the second one for future work. Since the confinement dynamics of $\mathcal{G}_{{\rm dark}}$ is a non-perturbative process, it is difficult to make quantitative predictions on the GW spectra that can be produced. The phase transition ultimately depends on the phase diagram of the theory, whose determination requires computationally expensive lattice computations. We can however have an order of magnitude estimate of the expected signal using effective models to parametrize the confining dynamics. As shown in~\cite{Helmboldt:2019pan}, in QCD-like theories different effective models give GW spectra whose peak amplitude might differ even by two orders of magnitudes between each other. Nevertheless they can give a useful indication of what kind of theories can produce a GW spectrum that is close enough to current and future experimental sensitivities, and that thus deserve further dedicated theoretical studies to allow for a more precise study of the phase transition and the associated GW signals. In this work we decide to focus on the linear sigma model description of a confining strong dynamics, which is defined in Eq.~\eqref{eq:strongly_interacting_generic} of Appendix~\ref{app:general_setup}. We nevertheless stress that our results should be taken as a preliminary indication for what the real GW spectrum could be, as recently also emphasized in~\cite{Helmboldt:2019pan,Croon:2019iuh}. Furthermore, it is important to stress that there are many sources of theoretical uncertainties in the computation of the GW spectrum. Among others, we can list the renormalization scale dependence, higher loop corrections, different ways to deal with the high temperature approximation, gauge dependence, and non-perturbative and nucleation corrections (see~\cite{Croon:2020cgk}).  It is also worthwhile to say that not all the uncertainties can be accurately quantified. For simplicity, we do not consider these additional uncertainties in our estimate of the GW spectrum since, as we have seen, we are describing non-perturbative processes that are, in any case, difficult to deal with.

As already mentioned in the Introduction, a well-known argument by Pisarski and Wilczek~\cite{Pisarski:1983ms} 
implies that $SU(N_d)$ gauge theories with $N_d\ge 3$ feature a first order phase transition if $N_F \ge 3$ light~\footnote{That is with a mass smaller than the new gauge theory confinement scale.} flavors are present.
Motivated by this result, we study here $SU(N_d)$ gauge theories with $N_F=3$ and $N_F=4$ flavors, considering both the situations in which the new confining phase transition happens at a scale below and above the EW critical temperature $T_{{\rm EW}}$. A general discussion of these models in the relaxion framework can be found in App.~\ref{app:explicit_models}. We focus on the regime in which the anomaly term dominates over the mass term. Since the only dependence of the Sigma model potential on the Higgs field comes through the mass term, we expect the Higgs dynamics to have only a very small impact on the phase transition. For simplicity, we will neglect such impact in our computation. We keep our discussion general, but to make it more concrete it is useful to take as an example the variations of the $L+N$ model shown in Fig.~\ref{fig:summary_models}, where $L$ and $N$ 
are new vector-like fermions in the fundamental of $SU(N_d)$ with the quantum number of the SM lepton doublet and of a total singlet, respectively. We see that the physics of the $N_F=3$ case is captured by the minimal $L+N$ model with dark confinement scale above the EW one (model A), as well as the physics of models in which we add $3$ singlets to the spectrum independently on the confinement scale (models B and D). Notice that the EW charged fermion $L$ cannot be too light to evade current experimental direct searches bounds, and we then assume it to have a mass larger than $\Lambda_{\rm EW}$. The physics of the $N_F=4$ dark flavors, on the contrary, is captured by the minimal model with two copies of $L$ fermions and a decoupled singlet (model C), and of models in which $L$ is decoupled and 4 light dark $N$ flavors are added to the spectrum (models B and D with $n_N = 4$). Experimental bounds on the minimal $L+N$ model have been considered in~\cite{Beauchesne:2017ukw,Antipin:2015jia,Barducci:2018yer}. Although important, these bounds allow for rather different spectra at the level of sigma model without spoiling the successful relaxation of the EW scale. 

\subsection{Computation of the gravitational wave spectrum}\label{sec:GW_comp}
We now quickly remind the reader how the computation of the spectrum of the stochastic GW background produced in a first order phase transition proceeds. Three contributions must be considered~\cite{Caprini:2018mtu}: the one from true vacuum bubble collisions, $\Omega_{\text{col}}\,h^{2}$, the one from the propagation of sound waves in the plasma, $\Omega_{\text{sw}}\,h^{2}$, and the one from magnetohydrodynamic turbulence effects, $\Omega_{\text{turb}}\,h^{2}$. A fourth contribution may be generated by quantum fluctuations~\cite{Mazumdar:2018dfl}, but since its impact is not well understood we will not consider it in the following. 

As mentioned, we describe the confining dynamics of the new strongly interacting sector through the linear sigma model described in Eq.~\eqref{eq:strongly_interacting_generic}. In this setup, the radial degree of freedom $\sigma$ interacts with the light particles in the plasma. We thus expect the interactions between the scalar shell and the plasma to be important, causing the behavior of the bubble to be a \textit{non-runaway} one~\cite{Caprini_2016}, that is the bubble walls do not keep accelerating until the bubbles collide. Since in this case most of the latent heat of the phase transition is transferred to the plasma in the form of sound waves and turbulence, in the following we will focus on the $\Omega_{\text{sw}}\,h^{2}$ and $\Omega_{\text{turb}}\,h^{2}$ contributions only. Their expressions can be written in a compact way as~\cite{Caprini_2016, Caprini:2019egz, Ellis_2019, Helmboldt:2019pan}
\be\label{eq:GW_spectr}
\Omega_i h^2 = a_i \,\left(\frac{\beta}{H}\right)^{-1}\, 
	\left(\frac{\kappa_{i}\,
		\alpha}{1 + \alpha}\right)^{b_i}\,
	\left(\frac{100}{g_{\star}}\right)^{1/3}\,
	v_{\text{wall} }\,S_{i}(f)\ ,
\ee
where $i =$ sw or turb for the sound waves and turbulence contributions, respectively. The details of the phase transition enter in the parameters $\alpha$, $\beta/H$ and $T_{n}$, which we will discuss in detail below. As for the other parameters, the numerical coefficients $a_i$ in the two cases are given by~\cite{Caprini_2016, Caprini:2019egz, Ellis_2019, Helmboldt:2019pan}
\be
a_{\rm sw} = 2.65 \times 	10^{-6}\ , ~~~~ a_{\rm turb} = 3.35 \times 	10^{-4}
\ee
and the exponents $b_{i}$ are
\be
b_{\rm sw} = 2\ , ~~~~ b_{\rm turb} = 3/2\ .
\ee
Again, $g_\star$ denotes the number of relativistic degrees of freedom and $H$ is the Hubble parameter, where both quantities are computed at the nucleation temperature. The quantity $v_{\rm wall}$ is the wall velocity. It is known that scenarios with large $v_{\rm wall}$ lead to stronger GW signals~\cite{Espinosa_2010,Caprini_2016,Croon:2019iuh,Helmboldt:2019pan}. While in the non-runway behavior the bubble walls stop their acceleration and reach a terminal velocity, this velocity might still be relativistic. Calculating the specific value   of $v_{\rm wall}$ is beyond the scope of the  present work. We will then concentrate on the case of highly relativistic bubbles, $v_{{\rm wall}}\sim 1$, since it is the most interesting regime from an observational point of view. Decreasing  $v_{\rm wall}$ down to $v_{{\rm wall}}\sim 0.75$ doesn't drastically modify the overall picture, see {\emph{e.g.}}~\cite{Helmboldt:2019pan}.
Finally, the coefficients $\kappa_i$ are the efficiencies of each process. The former 
is the efficiency to convert the latent heat into bulk motion, while the latter is the part that is converted into vorticity in the plasma. Numerical fits of $\kappa_{\rm sw}$ suggests the form~\cite{Espinosa_2010,Caprini_2016}
\be
\kappa_{\rm sw} \simeq \frac{\alpha}{0.73 + 0.083 \sqrt{\alpha} + \alpha}\ ,
\ee
for $v_{\rm wall} \sim 1$ where the quantity $\alpha$ is related to the strength of the transition and will be defined below. $\kappa_{\rm turb}$ is 
determined trough $\kappa_{\rm turb} = \epsilon \, \kappa_{\rm sw}$. Previous studies suggest the conservative value of $\epsilon= 5\times 10^{-2}$~\cite{Hindmarsh_2015,Caprini_2016,Ellis:2018mja}.

As regarding the spectral shapes $S_i(f)$, they read
\be
S_{\text{sw}}(f)  = 
	\frac{f^{3}}{f_{\text{sw}}^{3}}\,
	\left(\frac{7}{4 + 3
		\frac{f^{2}}{f_{\text{sw}}^{2}}
	}\right), ~~~~ 
S_{\text{turb}}(f) =
	\frac{f^{3}}{f_{\text{turb}}^{3}}\,
	\,\frac{\left(1+ \frac{f}{f_{\text{turb}}}\right)^{-\frac{11}{3}}}{1+ 8\,\pi\,f/ \tilde h
	}.
\ee
In the previous expression $\tilde h$ is the Hubble rate at $t_n$ redshifted to today,
\be
\tilde h  = 1.65 \times 10^{-5}\,\text{Hz}\,
\left(\frac{T_n}{100\,
	\text{GeV}}\right)\,\left(
\frac{g_\star}{100}\right)^{1/6}\ ,
\ee
while the peak frequencies in the two cases are given by similar expressions, 
\be\label{eq:fpeak}
f_i = f_i^0\, \frac{1}{v_{\rm wall}} \, \frac{\beta}{H}\, \frac{T_n}{100\,{\rm GeV}} \, \left( \frac{g_\star}{100} \right)^{1/6}\ ,
\ee
with $f_{\rm sw}^0 = 1.9 \times 10^{-5}$ Hz and $f_{\rm turb}^0 = 2.7 \times 10^{-5}$ Hz. 
It has been suggested in~\cite{Ellis:2018mja,Ellis_2019,Caprini:2019egz} that when $\beta/H\gg 1$, the 
sound wave and the turbulence contributions shown above \textit{overestimate} and \textit{underestimate} the signal, respectively. Following the suggestion in the same works, we modify $\Omega_{\text{sw}}\,h^{2}$ and $\Omega_{\text{turb}}\,h^{2}$ to
\be\label{eq:GW_spectr_fast}
\Omega^{\text{fast}}_{\text{sw}}\,h^{2} = \left(\tau_{\text{sw}}\,H\right)\,\Omega_{\text{sw}}\,h^{2}\ , ~~~ 
\Omega^{\text{fast}}_{\text{turb}}\,h^{2} =\left. \left(1-\tau_{\text{sw}}\,H\right)\,\Omega_{\text{turb}}\,h^{2}\right|_{\kappa_{\rm turb} = \kappa_{\rm sw}}\ ,
\ee
where 
$$
\tau_{\text{sw}}\,H=(8\pi)^{1/3}\,\frac{v_{\text{wall}}}{\overline{U}_{f}\,(\beta/H)}
$$ 
is related to the duration of sound waves and $\overline{U}_{f}$ is the root-mean-square four velocity of the plasma.

Let us now discuss the parameters directly related to the phase transition under consideration: the nucleation temperature $T_n$, the inverse time duration $\beta/H$ and the strength of the transition $\alpha$. The first two parameters are defined in terms of the nucleation rate~\cite{Coleman:1977py,Linde:1981zj}
\be
\Gamma(T) = \mathcal{A}_{(T)}\,e^{-S_{3}/T}\ ,
\ee
where $S_3$ is the Euclidian action computed at its bounce and $\mathcal{A}_{(T)}$ is a non-perturbative factor. The nucleation temperature is defined as the temperature at which one of the nucleated bubbles reaches a size comparable to the Hubble radius at that time, $\Gamma(T_n) \simeq H^{4}(T_n)$. Considering the process to take place in the radiation domination epoch, the above statement implies~\cite{Mazumdar:2018dfl,Breitbach_2019}
\be
	\frac{S_{3}}{T_n}\simeq 164.56
	- 2\log\left(\frac{g_*}{
		100}\right)
	- 4\log\left(\frac{T_n}{
		1\,\text{GeV}}\right)\,.
\ee
The exact numerical value of the first term on the right-hand side depends on the precise form of $\mathcal{A}_{(T)}$, for simplicity we assume $\mathcal{A}_{(T)} \simeq T^{4}$.  The inverse time duration $\beta/H$ is instead defined as the rate of change of $\Gamma$ at the nucleation time $t_{\text{nucl}}$ via~\cite{guo2020phase}
\be
	\beta \equiv \left.\frac{1}{\Gamma}\, 
	\frac{d\, \Gamma}{d\,t}\right|_{t_n}
	\Rightarrow
	\frac{\beta}{H} = \left.T\,\frac{d}{d T}
	\left(
	\frac{S_{3}}{T}\right)
	\right|_{T=T_n}.
\ee
The strength of the transition, $\alpha$, is instead not univocally defined. In the literature many definitions can be found~\cite{Espinosa_2010,Ellis_2019,Helmboldt:2019pan,Croon:2019iuh}. The main two are based on the latent heat and on the trance anomaly.  In the first one $\alpha$ is identified as the ratio between the latent heat of the transition and the energy density of the radiation in the plasma $\rho_{\rm rad}$. In the second one it is defined through the trace of the  the stress-energy tensor and $\rho_{\rm rad}$. From the practical point of view, both definitions can be summarized as
\be
\alpha = \frac{1}{\rho_{\rm rad}} \left.\left( \Delta V - n \, T \, \frac{\partial \Delta V}{\partial T}\right)\right|_{T = T_n}\ ,
\ee
where $\Delta V = V_{\text{eff}}(\sigma_{\text{false}}, T) - V_{\text{eff}}(\sigma_{\text{true}}, T)$ is the difference of the free energy density between the two phases and $n = 1$ ($1/4$) when $\alpha$ is defined via the latent heat (trace of the stress-energy tensor). 

\subsection{Gravitational wave spectrum: QCD-like case}\label{sec:QCD-like}
\begin{figure*}\centering
	\subfloat[$N_{F}= 3$, $T_{n}< T_{\rm EW}$]
	{\includegraphics[width=0.49\textwidth]{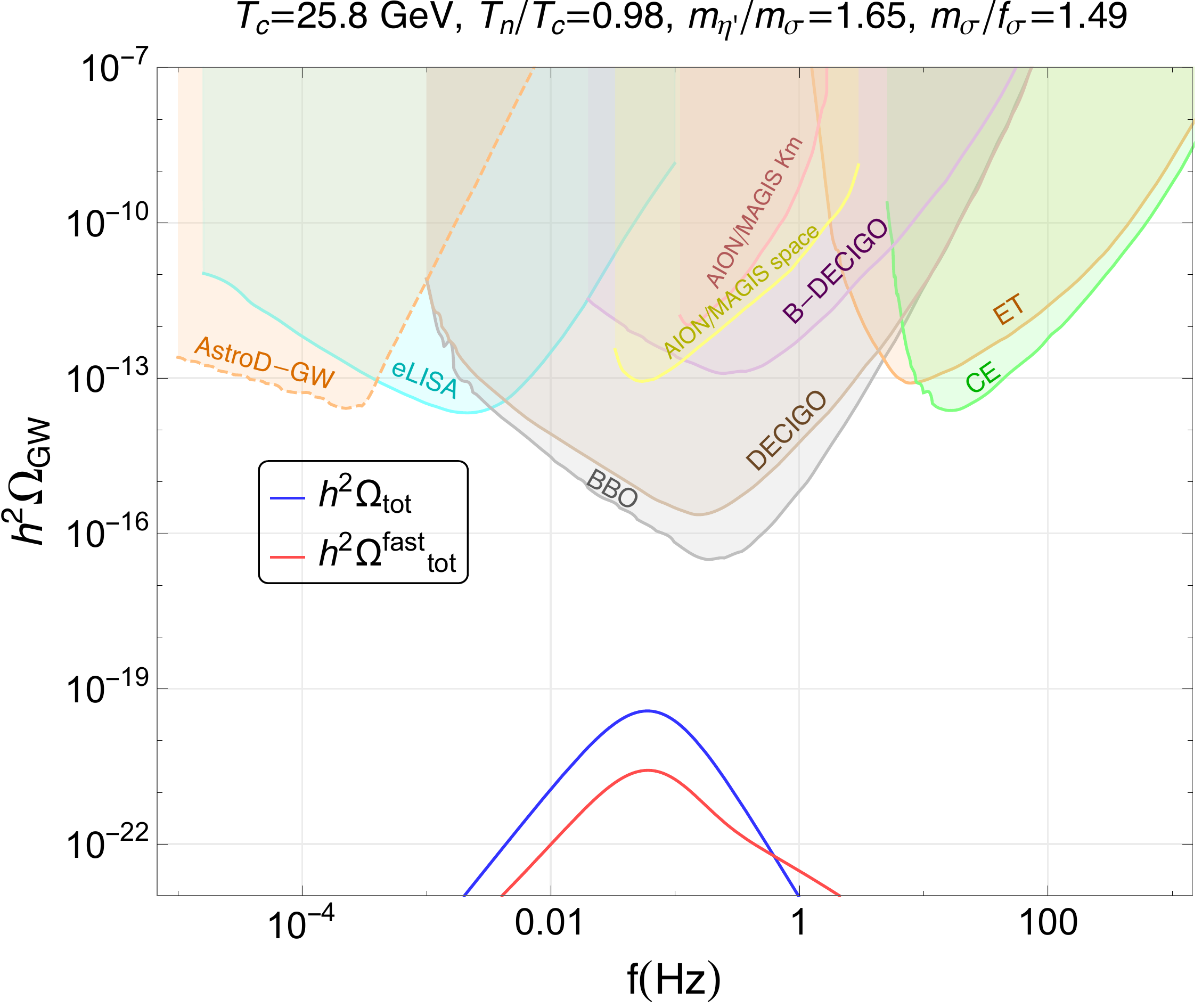}}
	\hfill
	\subfloat[$N_{F}= 3$, $T_{n}> T_{\rm EW}$]
	{\includegraphics[width=0.49\textwidth]{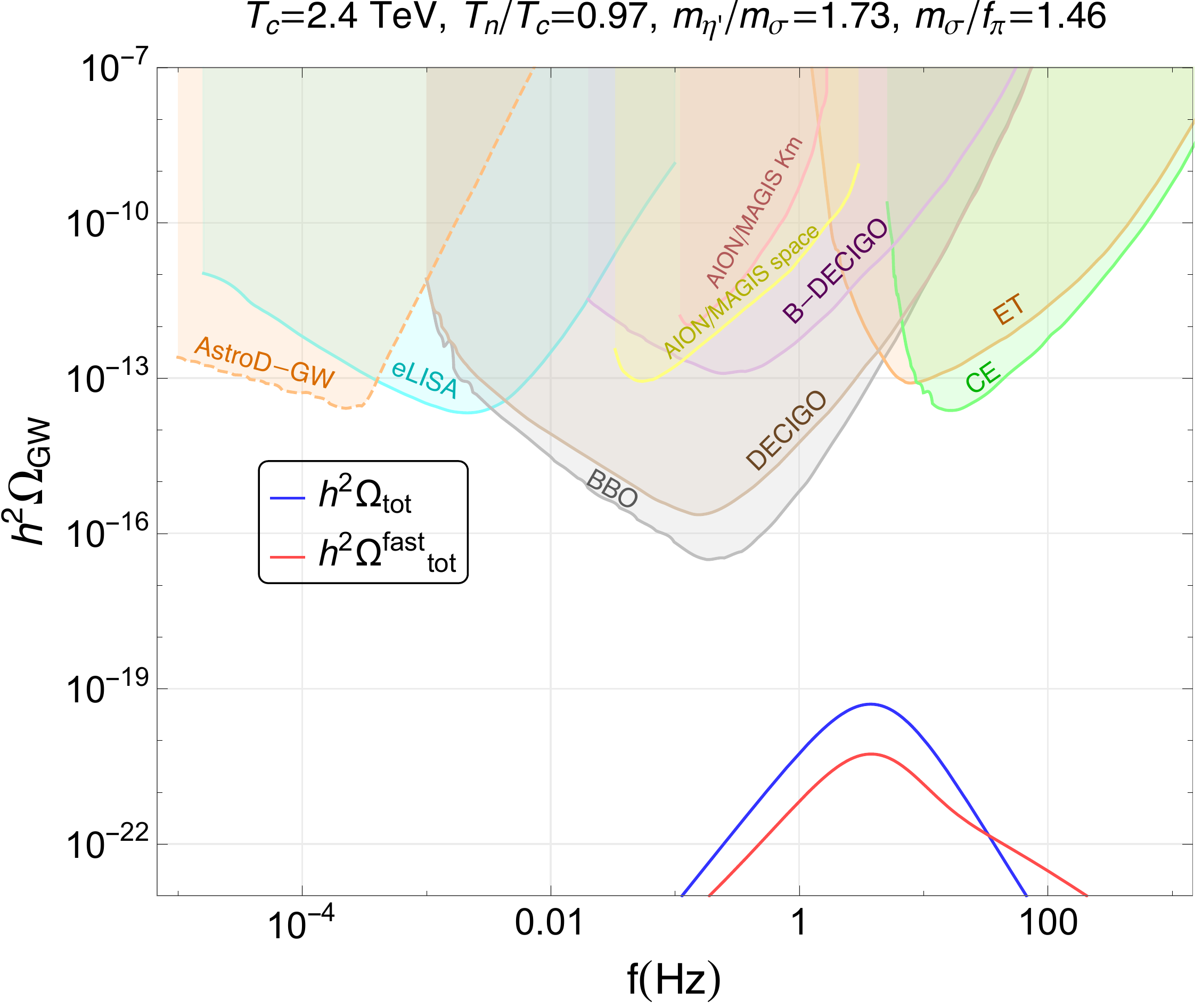}}
	\vspace*{.5cm}
	\subfloat[$N_{F}= 4$, $T_{n}< T_{\rm EW}$]
	{\includegraphics[width=0.49\textwidth]{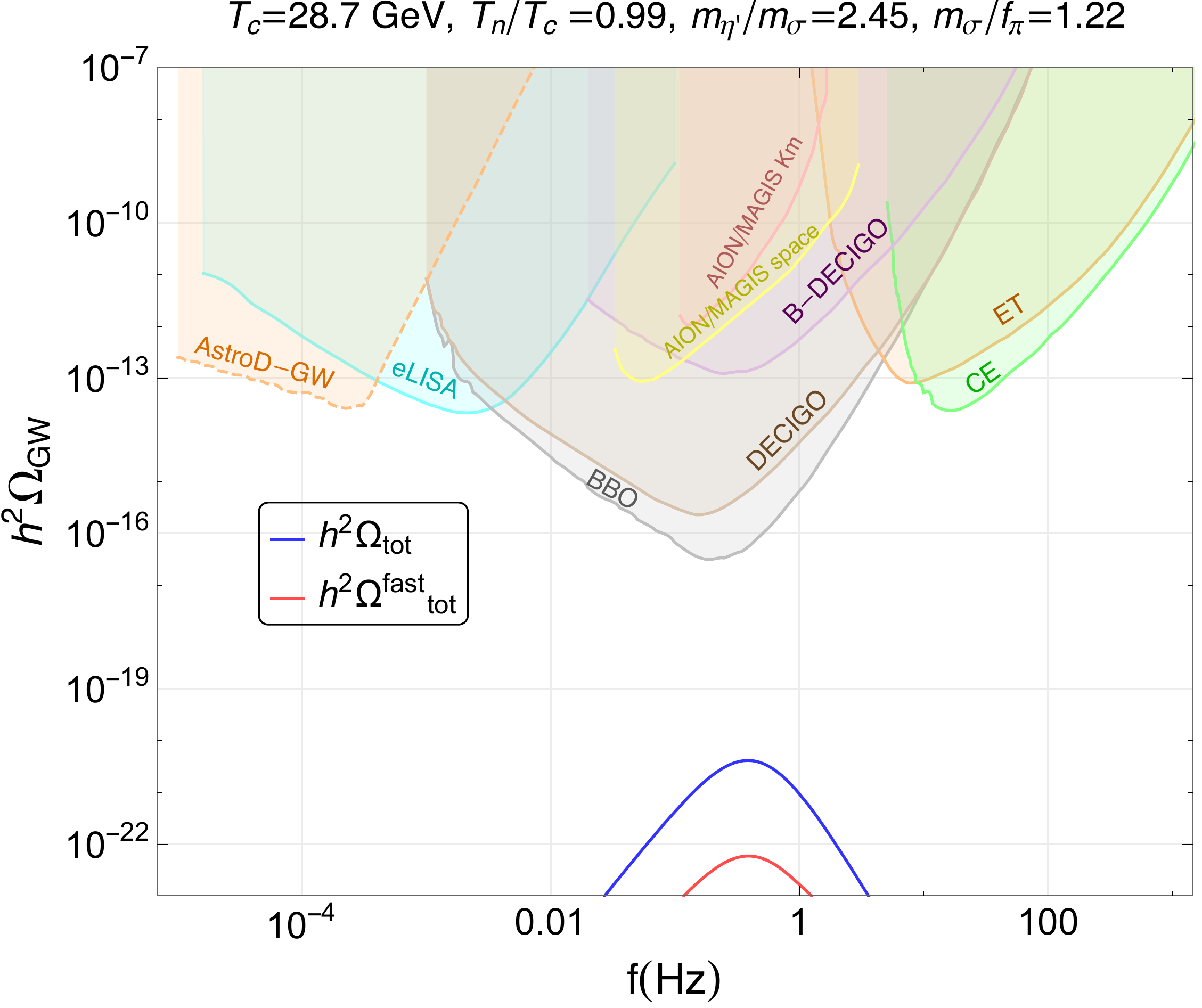}}
	\hfill
	\subfloat[$N_{F}= 4$, $T_{n}> T_{\rm EW}$]
	{\includegraphics[width=0.49\textwidth]{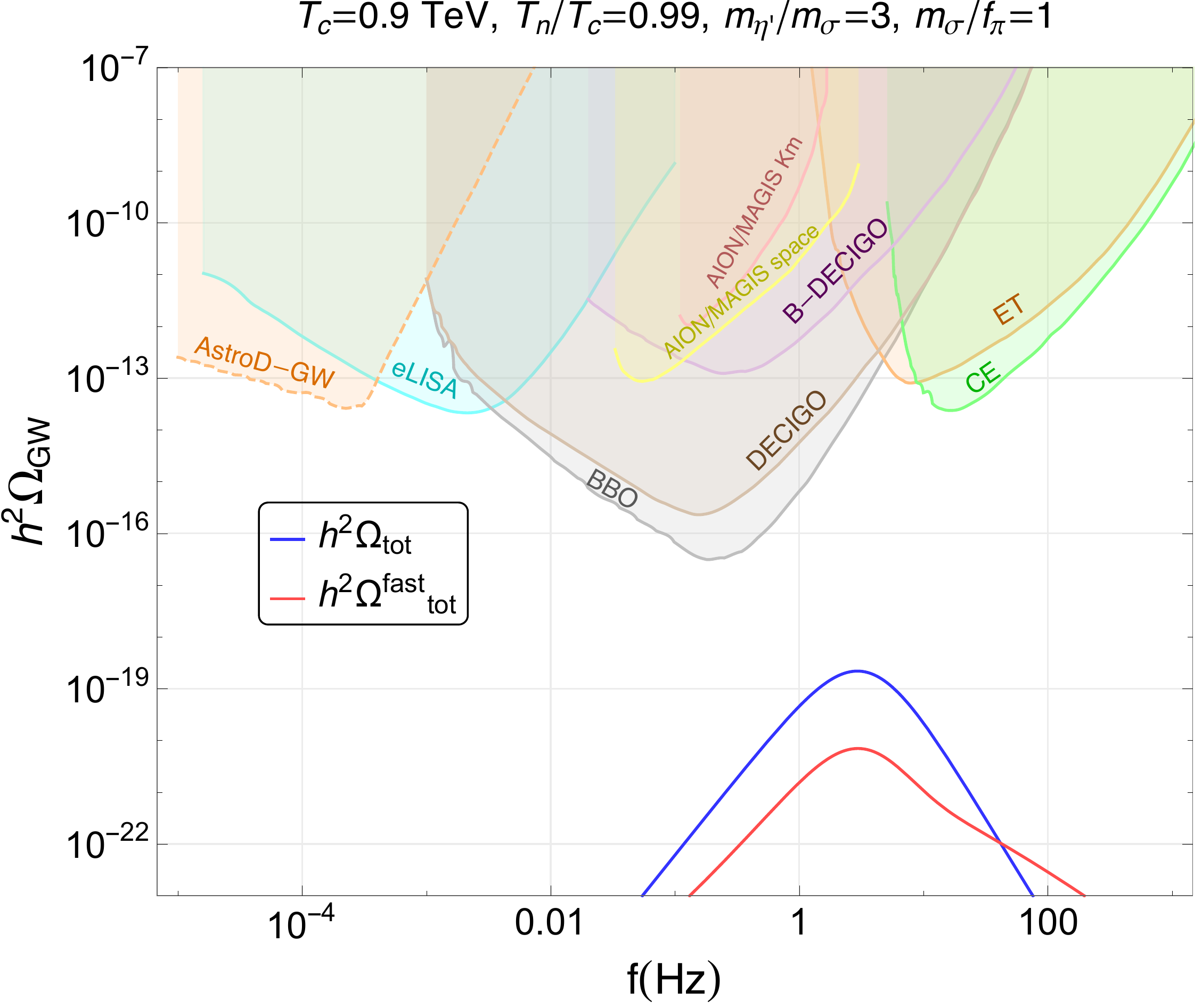}}
	\caption[]
	{Gravitational wave spectrum for a strongly interacting sector with a QCD-like spectrum, see the definition in Sec.~\ref{sec:QCD-like}. The parameters of the benchmark points shown are collected in Tab.~\ref{tab:lsm_pars_QCD}. We show the signal computed according to Eq.~\eqref{eq:GW_spectr} (blue line) and to Eq.~\eqref{eq:GW_spectr_fast} (red line).}
	\label{fig:QCDlike}
\end{figure*}

\begin{table}[t!]
\begin{center}
\scalebox{0.90}{
\begin{tabular}[t]{ c || c | c    }
        \multicolumn{3}{c}{{\bf{QCD-like case~~$N_F=3$}}}	   \vspace{0.3cm} \\
  & $T_n < T_{\rm EW}$ & $T_n > T_{\rm EW}$ \\ 
 \hline
 \hline
       $m_\Sigma^2\;$[$\text{GeV}^{2}$] 	&64 &  64  \\	
       $\lambda$        	& 1.5 & 1.5  \\	
       $\kappa$        	&4 & 4   \\	
       $\mu_\Sigma\;$[GeV]        	& $10^{2}$ & $10^{4}$  \\	    
\hline
	$f_{\pi}\;$[GeV]	& 24.7 &2.4$\times 10^3$\\
	$m_\sigma\;$[GeV] &36.9 & 3.4$\times 10^3$ \\
	$m_{\eta^\prime}\;$[GeV] &60.9 & 5.9$\times 10^3$\\
	$m_{S}\;$[GeV] & 60.8 &5.9$\times 10^3$\\
	$T_n\;$[GeV]	&25.2  & 2.3$\times 10^3$\\ 
\hline
	$\alpha$		& 0.00317& 0.0312\\
	$\beta / H$	& 11921.4& 7904.9               
\end{tabular}
}
\hspace{2cm}
\scalebox{0.90}{
\begin{tabular}[t]{ c || c | c    }
        \multicolumn{3}{c}{{\bf{QCD-like case~~$N_F=4$}}}	      \vspace{0.3cm} \\
  & $T_n < T_{\rm EW}$ & $T_n > T_{\rm EW}$ \\ 
 \hline
 \hline
       $m_\Sigma^2\;$[$\text{GeV}^{2}$]	& 676& $6.4\times 10^{5}$   \\	
       $\lambda$        	&  2&2   \\	
       $\kappa$        	&  4&3   \\	
       $\mu_\Sigma$        	& 9 &9   \\	    
\hline
	$f_{\pi}\;$[GeV]	&30 &1.1$\times 10^3$\\
	$m_\sigma\;$[GeV] & 36.8 & 1.1$\times 10^3$\\
	$m_{\eta^\prime}\;$[GeV] & 90.1 &3.4$\times 10^3$\\
	$m_{S}\;$[GeV] & 76.5 &2.8$\times 10^3$\\
	$T_n\;$[GeV]	& 28.6 &0.9$\times 10^3$\\ 
\hline
	$\alpha$		&0.00283&0.0054 \\
	$\beta / H$	&66894.1&15694.5               
\end{tabular}
}
\caption{Values of the linear sigma model parameters, physical meson masses and relevant quantities entering the GW spectrum calculation of the $N_F=3$ and $N_F=4$ for the QCD-like cases. }
\label{tab:lsm_pars_QCD}
\end{center}
\end{table}
We now compute the GW spectrum using the linear sigma model defined in App.~\ref{app:general_setup}
with the formalism of Sec.~\ref{sec:GW_comp}. We refer the reader to the Appendix for a definition of the notation we use for the linear sigma model. We focus here on the situation in which the masses of the $\sigma$ and $\eta'$ mesons satisfy $m_{\sigma, \eta'} \gtrsim f_{\pi}$, where $f_{\pi}$ is the pion decay constant. Since this is the well-known spectrum of QCD, we use the term ``QCD-like'' to refer to this case. We will study in Sec.~\ref{sec:nonQCD} a different physical spectrum. In our computation we consider the chiral limit, in which the mass of the lightest fermions is much smaller than the confinement scale $\Lambda_d$ and we focus on the $N_F=3$ and $N_F=4$ cases. We present our results for two specific values of $T_n$: one where $T_n < T_{{\rm EW}}$ and one where $T_n > T_{\rm EW}$, thus capturing the different possibilities discussed in Sec.~\ref{sec:evolution_after_reheating}. In particular we fix the values of the linear sigma model parameters for the two cases as reported in Tab.~\ref{tab:lsm_pars_QCD}, where we also indicate the values of the chiral symmetry breaking VEV, the relevant mesons masses, the nucleation temperature $T_n$ and of the main parameters entering the GW spectrum computation. In choosing these benchmark point we have scanned the linear sigma models parameters trying to maximize the GW spectrum for both nucleation temperatures. In the lack of analytical expressions for $\alpha$ and $\beta/H$ this represents a challenging numerical process.
As already stressed in~\cite{Helmboldt:2019pan} it's interesting to notice that large values of $\beta/H$ are found. This is to be contrasted with the naive estimate
\be
\frac{\beta}{H} \simeq 4 \log \frac{M_{{\rm Pl}}}{T_n} \simeq {\cal O}(100) \ .
\ee
Since for the sound waves and  turbulence contribution the GW amplitude decreases linearly  with $\beta/H$, see Eq.~\eqref{eq:GW_spectr}, while the peak frequency increases linearly with it, see Eq.~\eqref{eq:fpeak}, the direct computation of $\beta/H$ through an explicit, although effective, model has a strong consequence for the observability of the GW spectra. All together our results for the QCD-like case are shown in Fig.~\ref{fig:QCDlike} for the $N_F=3$  and $N_F=4$ models in the upper and lower panels respectively. There,   the left and right panels correspond to the $T_n<T_{\rm EW}$ and $T_n>T_{\rm EW}$ cases. In all plots the blue line shows the spectrum computed using Eq.~\eqref{eq:GW_spectr}, while the red line is  computed using the modification presented in Eq.~\eqref{eq:GW_spectr_fast}, which explicitly show the suppression factor due to the decrease of the sound wave contribution.  The situation depicted in the left panels can be realized, in the relaxation framework we are considering, when $T_n < T_{{\rm RH}} < T_{{\rm EW}}$ or when $T_n < T_{{\rm EW}} < T_{{\rm RH}}$. As shown in Sec.~\ref{sec:evolution_after_reheating} in the first case large regions of parameter space end up with large variations on $v_{{\rm EW}}$ from the second relaxation phase, leaving thus $T_{{\rm EW}} < T_{{\rm RH}}$ as the preferred case~\footnote{We remind the reader that there are however choices of parameters for which the variation $\delta v/v$ is smaller than one even when $T_{{\rm RH}} < T_{{\rm EW}}$, see Fig.~\ref{fig:vev_relative_difference_1}.}.
The situation of the right panels refers instead to the case $T_{{\rm EW}} < T_n < T_{{\rm RH}}$, for which we showed that large regions of parameter space are compatible with a small variation of the EW VEV during the second relaxation phase. 
In all the figures the colored regions represent the sensitivities of future interferometer experiments. In particular we show the projected reach from AstroD-GW~\cite{NI_2013,Kuroda_2015}, eLISA~\cite{amaroseoane2012elisa,Kuroda_2015}, BBO~\cite{Ni_2016,Kuroda_2015}, DECIGO~\cite{Ni_2016,Kuroda_2015}, B-DECIGO~\cite{Ni_2016,Kuroda_2015}, AION~\cite{Badurina_2020}, MAGIS~\cite{graham2017midband}, ET~\cite{Maggiore_2020} and CE~\cite{reitze2019cosmic}.  As we see from Fig.~\ref{fig:QCDlike}, both for $N_F=3$ and $N_F=4$ cases the GW signal from the dark phase transition is 
a few orders of magnitude below the region that can be probed in future experiments, in agreement with previous results obtained in similar frameworks~\cite{Helmboldt:2019pan, Croon:2019iuh}.
We stress again, however, that these results must be interpreted as an order-of-magnitude estimate since, as shown in~\cite{Helmboldt:2019pan}, different effective models for the strong sector confining dynamics can give results that might differ even by two orders of magnitude for the amplitude of the signal and that might then fall on the edge of detectability. Notice also that changing $N_F$ does not dramatically change the region in which the signal falls, making thus difficult the identification of the underlying model in case a signal is detected. We conclude by emphasizing once more that, according to the discussion in Appendix~\ref{app:explicit_models}, the strongly interacting models generating the back-reaction potential in the relaxion mechanism do not suffer from very strong experimental limits. In particular, it is not difficult to obtain in such models the spectrum used in this section. 

\subsection{Gravitational wave spectrum: non QCD-like case} \label{sec:nonQCD}
\begin{table}[t!]
\begin{center}
\scalebox{0.90}{
\begin{tabular}[t]{ c || c | c    }
        \multicolumn{3}{c}{{\bf{Non QCD-like case~~$N_F=3$}}}	   \vspace{0.3cm} \\
  & $T_n < T_{\rm EW}$ & $T_n > T_{\rm EW}$ \\ 
 \hline
 \hline
       $m_\Sigma^2\;$[$\text{GeV}^{2}$] 	&1 & 1   \\	
       $\lambda$        	&0.01  &0.01   \\	
       $\kappa$        	& 0.1 & 0.1  \\	
       $\mu_\Sigma\;$[GeV]        	&3.5  & 150   \\	    
\hline
	$f_{\pi}\;$[GeV]	&54.4  &2.3$\times 10^3$\\
	$m_\sigma\;$[GeV] & 9.86  &0.4$\times 10^3$\\
	$m_{\eta^\prime}\;$[GeV] &16.90  &0.7$\times 10^3$\\
	$m_{S}\;$[GeV] & 18.4 &0.8$\times 10^3$\\
	$T_n\;$[GeV]	&19.9  &0.8$\times 10^3$\\ 
\hline
	$\alpha$     &0.00336 & 0.00348\\
	$\beta / H$	&1166.25 & 907.5               
\end{tabular}
}
\hspace{2cm}
\scalebox{0.90}{
\begin{tabular}[t]{ c || c | c    }
        \multicolumn{3}{c}{{\bf{Non QCD-like case~~$N_F=4$}}}	      \vspace{0.3cm} \\
  & $T_n < T_{\rm EW}$ & $T_n > T_{\rm EW}$ \\ 
 \hline
 \hline
       $m_\Sigma^2\;$[$\text{GeV}^{2}$] 	&25 & $4.9\times 10^{5}$   \\	
       $\lambda$        	& 0.3 & 2  \\	
       $\kappa$        	& 0.4 & 3  \\	
       $\mu_\Sigma$        	&1.56  & 10.58   \\	    
\hline
	$f_{\pi}\;$[GeV]	&50  & 2.1$\times 10^3$\\
	$m_\sigma\;$[GeV] &7  &0.9$\times 10^3$ \\
	$m_{\eta^\prime}\;$[GeV] &62.4  & 7$\times 10^3$\\
	$m_{S}\;$[GeV] & 49.5 &5.6$\times 10^3$\\
	$T_n\;$[GeV]	&9  &0.9$\times 10^3$\\ 
\hline
	$\alpha$  &0.09378 & 0.02136\\
	$\beta / H$	& 1483.2 & 1828.3              
\end{tabular}
}
\caption{Values of the linear sigma model parameters, physical meson masses and relevant quantities entering the GW spectrum calculation of the $N_F=3$ and $N_F=4$ for the non QCD-like cases. }
\label{tab:lsm_pars_non_QCD}
\end{center}
\end{table}
We now turn to the discussion of more exotic strongly interacting models, {\it i.e.} models in which, unlike the case of QCD, the theory spectrum features $m_{\sigma} \lesssim f_{\pi}$. The behavior of gauge theories with different number of flavors has been studied on the lattice in a certain number of situations, with the surprising result that a non-QCD behavior in which the $\sigma$ meson is lighter than expected might emerge. In particular, light composite $\sigma$ scalars have been found in an SU(3) gauge theory with 8 flavors in the fundamental~\cite{Aoki:2014oha, Aoki:2016wnc, Appelquist:2016viq, Appelquist:2018yqe}, with 2 flavors in the symmetric representation~\cite{Fodor:2014pqa, Fodor:2015vwa, Fodor:2016pls} and with 4 light and 8 heavy flavors~\cite{Brower:2015owo}. They also appear in SU(2) theories with one adjoint flavor~\cite{Athenodorou:2015fda}. The behavior seems to be quite generic and is typically associated with gauge theories near to their conformal limit. In the case of $N_F=3$ and $N_F=4$, this is expected to happen when the number of colors is larger than $4$ and the fermions transform in the antisymmetric representation~\cite{Ryttov:2007sr}. 

\begin{figure*}\centering
	\subfloat[$N_{F}= 3$, $T_n< T_{EW}$]
	{\includegraphics[width=0.49\textwidth]{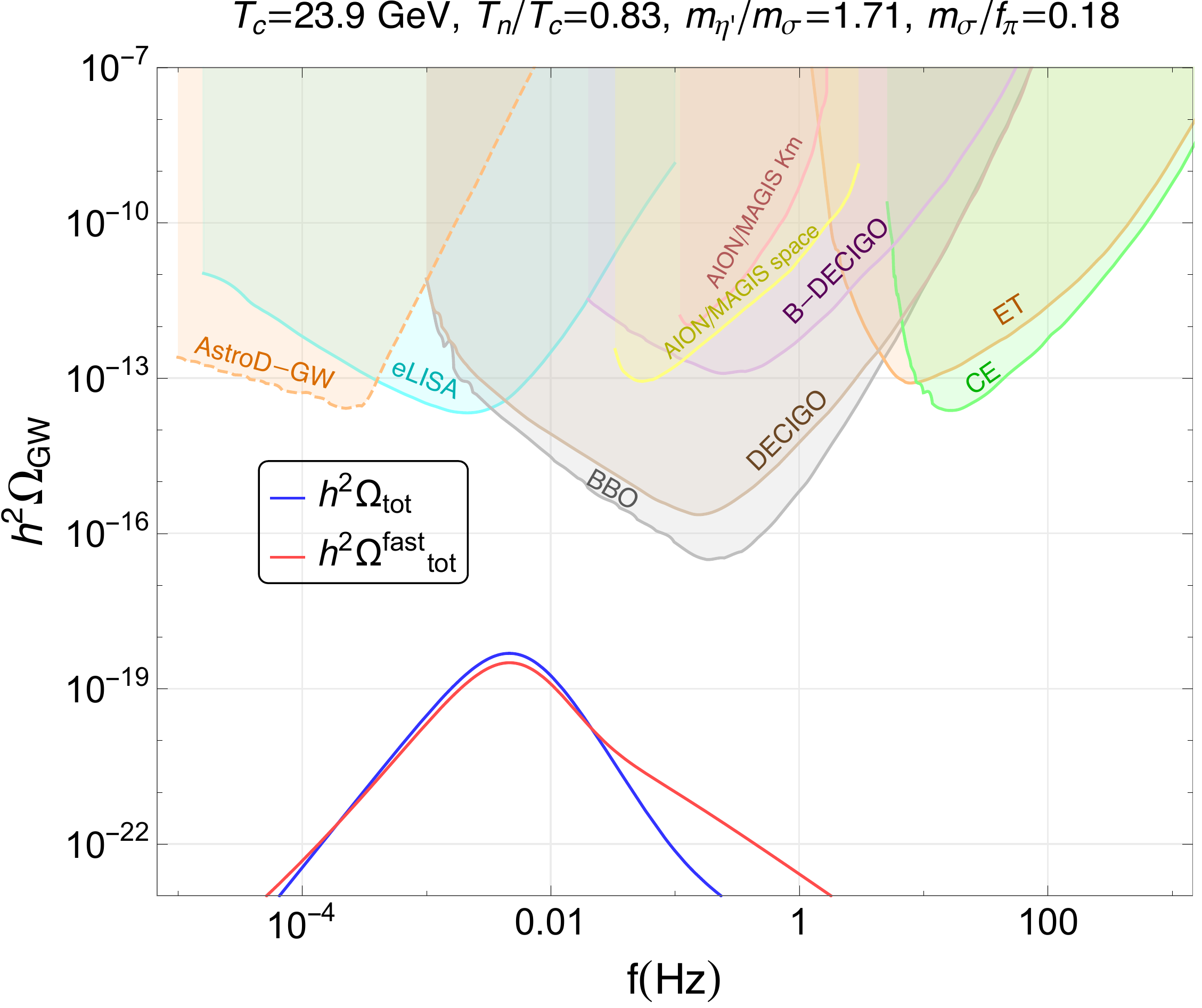}}
	\hfill
	\subfloat[$N_{F}= 3$, $T_n> T_{EW}$]
	{\includegraphics[width=0.49\textwidth]{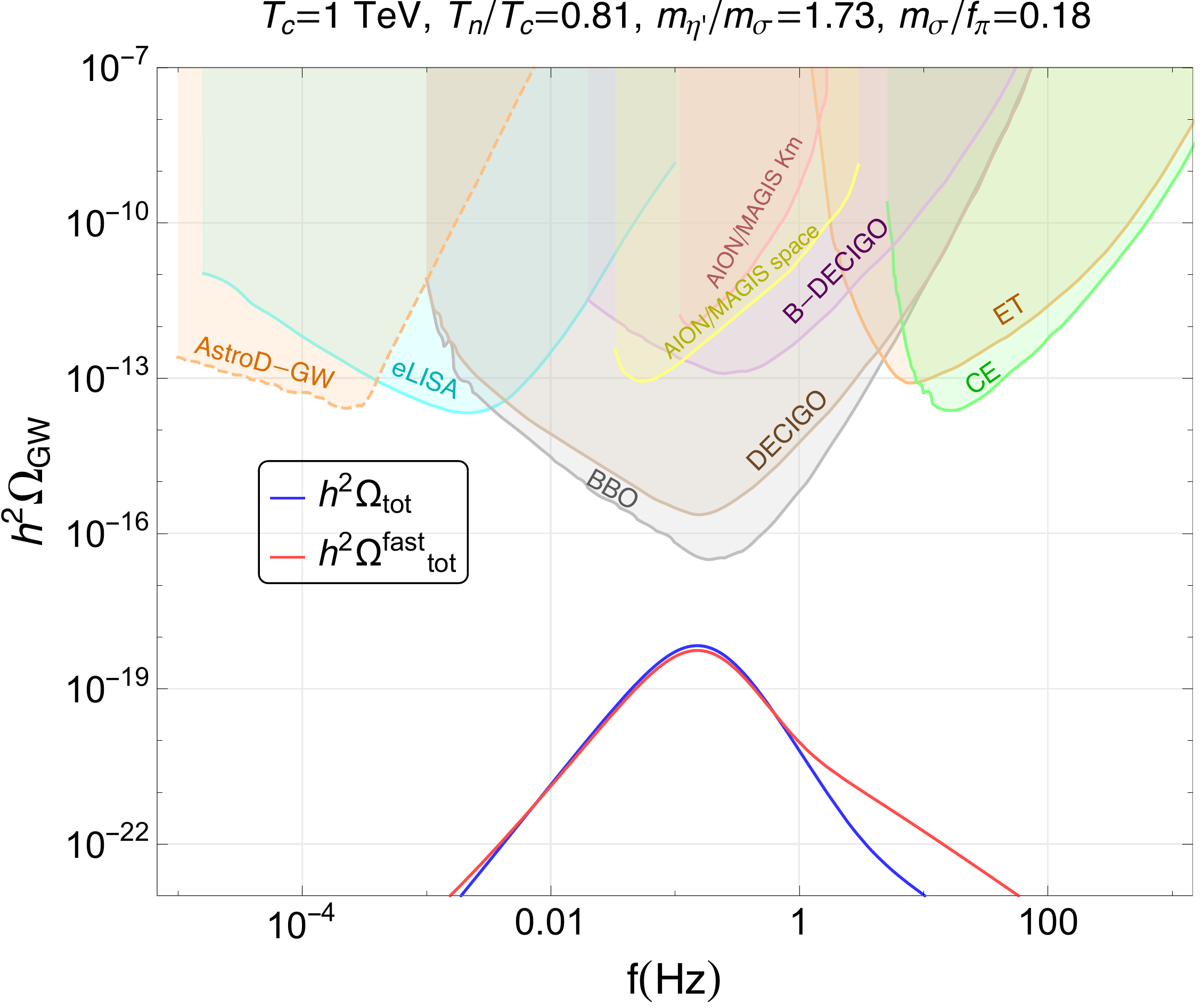}}
	\vspace*{.5cm}
	\subfloat[$N_{F}= 4$, $T_n< T_{EW}$]
	{\includegraphics[width=0.49\textwidth]{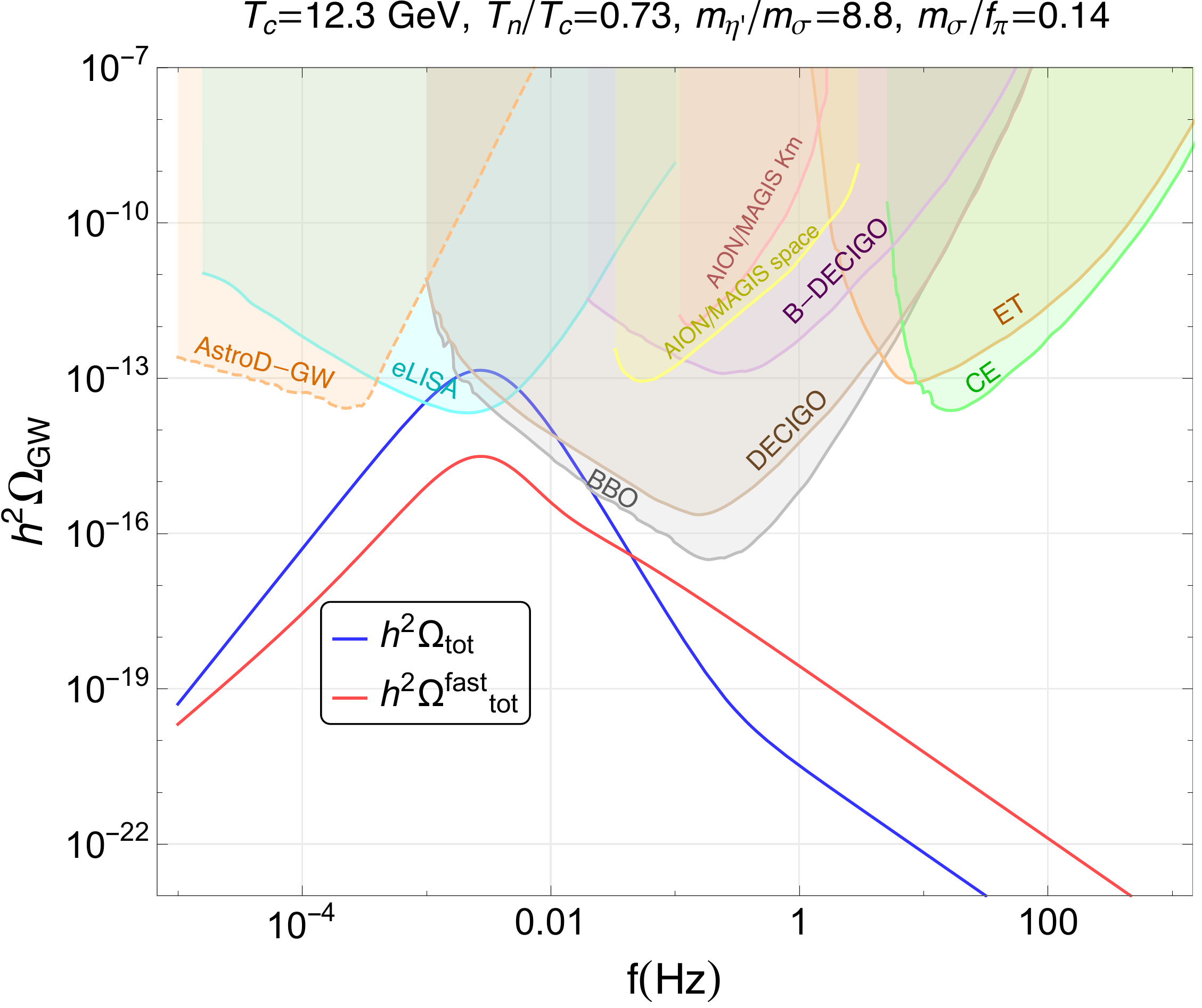}}
	\hfill
	\subfloat[$N_{F}= 4$, $T_n> T_{EW}$]
	{\includegraphics[width=0.49\textwidth]{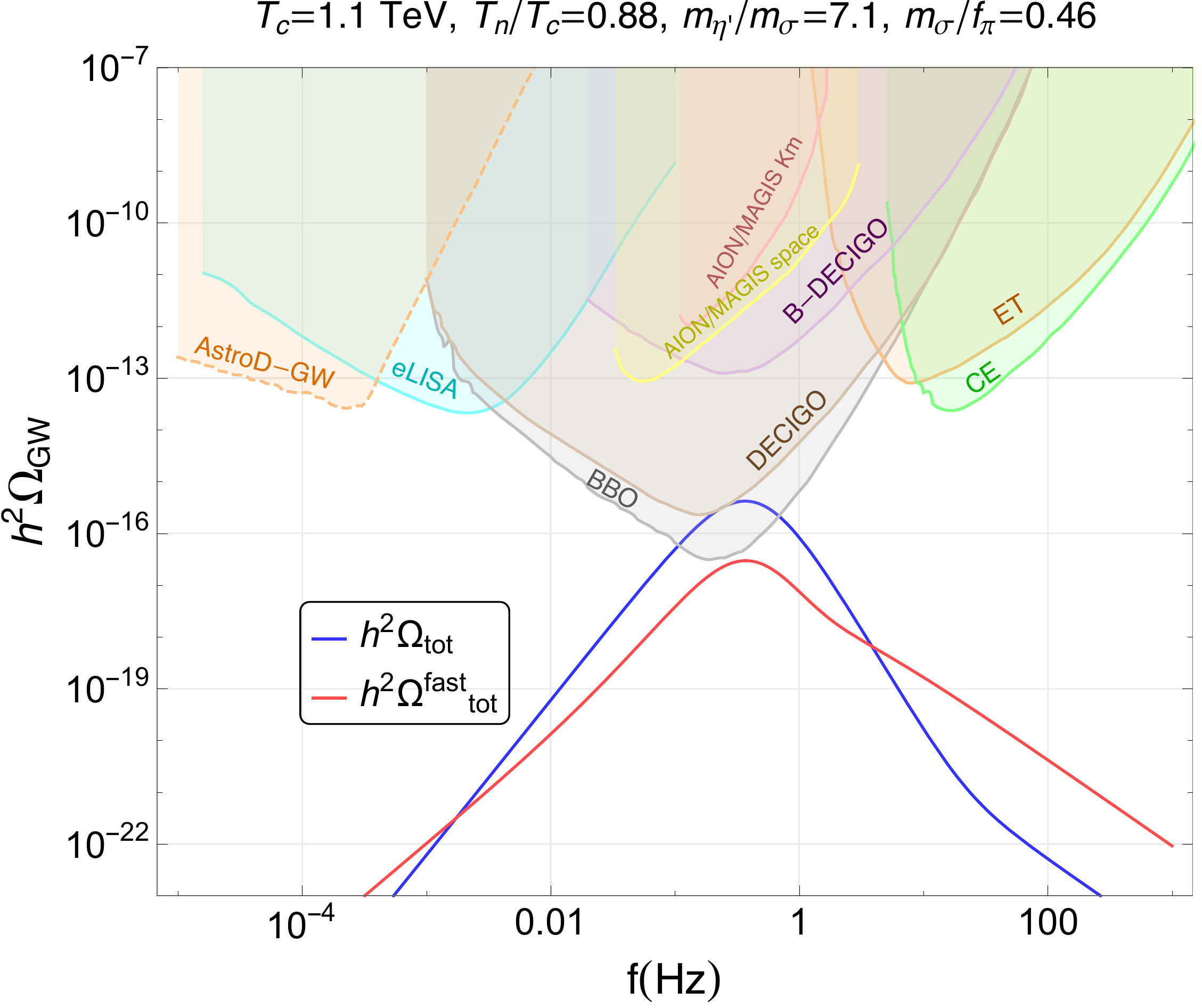}}
	\caption[]
	{Same as in Fig.~\ref{fig:QCDlike} but for the non QCD-like case defined in Sec.~\ref{sec:nonQCD}. The parameters of the benchmarks shown are collected in Tab.~\ref{tab:lsm_pars_non_QCD}. }
	\label{fig:nonQCDlike}
\end{figure*}

In all the cases mentioned above, the $\sigma$ meson is found to be roughly degenerate with the pions, at least in the limit of large chiral symmetry breaking. Such behavior can be captured by a sigma model, as shown in~\cite{Appelquist:2018tyt}. The limit of small chiral symmetry breaking is however more difficult to describe on the lattice. In the absence of conclusive data, we will  suppose that the physics is correctly captured by a sigma model, at least in a first approximation. In~\cite{Meurice:2017zng} such sigma model has been extended to include the effects of the $\eta'$ mass. An interesting result emerging from the analysis is that the degeneracy between $m_\sigma$ and $m_\pi$ can be explained by an approximate cancellation between the VEV and $m_{\eta'}$. If this happens in the chiral limit, an immediate consequence is that typically $m_\sigma \lesssim$ VEV, and thus $m_\sigma \lesssim f_{\pi}$. This is the situation studied in this section. 
  Our results are shown in Fig.~\ref{fig:nonQCDlike}, where the color codes are the same as in Fig.~\ref{fig:QCDlike}. 
  
  As we see, allowing for $m_\sigma \lesssim f_{\pi}$ allows to boost the GW signal amplitude, in agreement with the results of~\cite{Croon:2019iuh}, where detectable GW spectra where found for large values of $m_{\eta^\prime}/m_{\sigma}$. Altogether we find that, while in the $N_F=3$ case also in the case of non QCD-like theory the predicted GW signal lies well below the reach of future experiments, in the $N_F=4$ case, the signal could be potentially detected both for the $T_c < T_{\rm EW}$ and $T_c > T_{\rm EW}$ cases, in the frequency range $10^{-3}\;{\rm Hz}-1\;{\rm Hz}$. Numerically, we have also explicitly checked that other than a large 
  $m_{\eta^\prime}/m_\sigma$, also the condition of a small $m_\sigma/f_\pi$ should be satisfied to enhance the signal towards the reach of future experiments. If any one of these two conditions fails to be satisfied, the signal typically lies well below the region of future detectability.

\section{Conclusions}\label{sec:concl}
In this paper we have considered the framework in which the EW scale is stabilized through the {\emph{relaxation}} mechanism. We have assumed that this happens during inflation and that the back-reaction potential needed to stop the relaxion evolution is generated by new vector-like fermions charged under a new strongly interacting $SU(N_d)$ gauge group.
We have focused on a configuration where the reheating temperature is above the confinement scale of the new strong dynamics. This causes the restoration of a deconfined phase after inflation, the disappearance of the 
back-reaction potential and the presence of a second relaxation phase during the early Universe thermal evolution. This second relaxation phase can in principle completely spoil the solution of the hierarchy problem. 
Once the temperature of the plasma drops again below the confinement scale of the new strong dynamics, the barrier again forms  and the relaxation finally stops its evolution. Crucially, the phase transition between the confined and unconfined dynamics might be strongly first order and can then produce a stochastic GW background, that  can be detected at present and future interferometer experiments. 

We have studied the relaxation evolution during the second relaxation phase, finding analytical solutions for its equation of motions for various ranges of temperatures. In particular, we have shown that, depending on the model parameters and on the hierarchy between the nucleation, reheating and EW phase transition temperatures, there are ample regions in parameter space where the second relaxation phase does not spoil the solution to the hierarchy problem. Such regions are large when $T_n < T_{\rm EW} < T_{\rm RH}$ and $T_{\rm EW} < T_n < T_{\rm RH}$, but are typically small or inexistent when $T_n < T_{\rm RH} < T_{\rm EW}$. 

We have then studied the GW signal that can be generated during the confining phase transition that ends the second relaxation stage, considering $SU(N_d)$ gauge theories with 3 and 4 light flavors present in the spectrum. To quantitatively describe the strong dynamics we have employed a linear sigma model, considering both QCD-like spectra, in which the $\sigma$ meson is heavier than the symmetry breaking scale, and non QCD-like spectra, in which the $\sigma$ meson can be lighter than it. The latter behavior may emerge in theories close to their conformal window, although additional lattice studies are needed to establish whether this is the case or not. While in the first case we find that the predicted signals lie below the present and future experimental sensitivities, in the case of non-QCD like spectra signals close and within the experimental reach can be obtained for the $N_F=4$ case. We however observe that, even if a GW signal will be detected in the future, the reconstruction of the underlying model will in general be challenging. On the one hand, as we have shown, there is little difference in the signal shapes expected for $N_F = 3$ and $N_F=4$ cases we have analyzed. On the other hand, many different models of strongly interacting vector-like fermions can give rise to the same relaxion back-reaction potential and can be described through the same linear sigma model studied in this paper.  We finally stress that all results obtained in this work by describing the dynamics of a strongly interacting theory through effective models suffer by large uncertainties, that can affect the peaks positions and heights of the predicted GW spectra~\cite{Helmboldt:2019pan}. Nevertheless we believe that is of paramount interest that BSM physics that can offer a solution to the hierarchy problem through the relaxion mechanism might generate a GW signal in the range of detectability of future experiments, and this makes even more important a thorough study of such theories through first principle calculations.

\acknowledgments

We are indebted to Djuna Croon and Rachel Houtz for clarifications on their related works, and to Bithika Jain for collaboration in the early stages of the project. DB~thanks Andrea Barducci for discussions.
This project has received funding from the European Union's Horizon 2020 research and innovation programme under the Marie Sk\l odowska-Curie grant agreement No 690575. DB thanks the Universidade de S\~{a}o Paulo for hospitality  while part of this work was carried out.  EB acknowledges financial support from FAPESP under contracts 2015/25884-4 and 2019/15149-6, and is indebted to the Theoretical Particle Physics and Cosmology group at King's College London for hospitality. MA thanks the Brazilian National Council for Scientific and Technological Development (CNPq) for his PhD scholarship - Process 140884/2017-3.

\appendix
\section{Strongly interacting models for the relaxation of the EW scale}\label{sec:strong_int_models}

We collect in this Appendix some useful formulas regarding strongly interacting vector-fermion models and their vacuum energy. We start with some general results and then specialize to the models used in the relaxion framework. 

\subsection{General setup}\label{app:general_setup}

Consider $N_F$ vector-like fermions $\Psi$ charged under a new confining group, which for simplicity we take to be $SU(N_d)$. We will assume a situation similar to what happens in QCD, namely that there is a small explicit breaking of the chiral symmetry $SU(N_F)_L \times SU(N_F)_R$ due to non vanishing fermion masses. Moreover, we take the axial part of the global flavor group to be anomalous. The Lagrangian we consider is
\be
{\cal L} = i \bar{\Psi}  \slashed{\partial} \Psi  -\overline{\Psi}_L {\cal M}  \Psi_R + \frac{\theta_0}{32\pi^2} D_{\mu\nu}\tilde{D^{\mu\nu}} + h.c.\ ,
\ee
where ${\cal M}$ is the $n\times n$ mass matrix of the vector-like fermions $\Psi$ and $D_{\mu\nu}$ is the field strength tensor of the $SU(N_d)$ gauge group, with $\tilde D_{\mu\nu}$ its dual. Using a $SU(N_F)_L \times SU(N_F)_R$ transformation~\footnote{Note that the phases can be factorized using a transformation generated by the diagonal elements of the two groups.} the mass matrix can always be put in the form
\be
{\cal M} \to e^{i \varphi_M/N_F} {\rm diag}( M_1 , \dots , M_{N_F}) \equiv e^{i \varphi_M/N_F} {\cal M}_D \ ,
\ee
where $\varphi_M = {\rm arg}\det{\cal M}$ and $M_i \ge 0$. We use this basis in the following. We can write the low energy theory by using the following transformations and spurions under an axial transformation 
\be
\Psi_{L,R} \to e^{\mp i \alpha} \Psi_{L,R}\ , ~~ M \to e^{-2 i \alpha} M\ , ~~ \theta_0 \to \theta_0 - 2 \,N_F \, \alpha \ .
\ee
We collect the low energy degrees of freedom in a matrix
\be
\Sigma = `` \overline{\Psi}_L   \Psi_R '' =\left(\frac{v + \sigma}{\sqrt{2 N_F}} + S^a T^a \right) \, {\cal U} \ ,
\ee
where $\sigma$ is the radial degree of freedom, $T^a$ are the $SU(N)$ generator, $S^a$ are CP-even scalars and ${\cal U}$ is the matrix of the NBGs, including the dark $\eta'$. Notice that we denote the VEV with $v$, as opposed to the EW VEV, which has been called $v_{{\rm EW}}$. We write $\mathcal{U}$ explicitly as
\be
\mathcal{U} = \mathrm{exp}\left[i \left( \frac{\eta'}{v} + \frac{\pi^a T^a}{v/\sqrt{2 N_F}} \right)\right]\ .
\ee
Following~\cite{Meurice:2017zng} we define the pion decay constant in a theory with $N_F$ flavors as
\be
f_\pi = \sqrt{\frac{2}{N_F}} v\ .
\ee
The most general potential invariant under $SU(N_F)_L \times SU(N_F)_R \times U(1)_V \times U(1)_A$ is given by
\be\label{eq:strongly_interacting_generic}
V  = - m_\Sigma^2 \langle \Sigma^\dag \Sigma \rangle + \frac{\lambda}{2} \langle \Sigma^\dag \Sigma \rangle^2 +  \frac{\kappa}{2} \langle (\Sigma^\dag \Sigma)^2 \rangle  - B\, \langle {\cal M} \Sigma \rangle + \mu_\Sigma\, e^{- i \theta} \det\Sigma + h.c.
\ee
where $\langle \cdot \rangle$ denotes the trace and we have used an axial transformation to put the phase dependence in the anomaly term, $\theta = \theta_0 + \varphi_M$. 

Let us start with the computation of the vacuum energy, since it is essential to write the relaxion potential in Eq.~\eqref{eq:potential}. In doing this we ignore the heavy degrees of freedom $\sigma$ and $S^a$ in Eq.~\eqref{eq:strongly_interacting_generic}. Using the results of~\cite{Gasser:1984gg}, working in the basis in which the fermion mass matrix is diagonal forces the $\Sigma$ matrix in the vacuum to be diagonal. We write it as
\be\label{eq:vacuum_Sigma_nf}
\Sigma_0 \equiv \langle \Sigma \rangle = \frac{v}{\sqrt{2 N_F}}{\rm diag} (e^{i \theta_1}, \dots , e^{i \theta_{N_F}})\ ,
\ee
which gives a vacuum energy
\be\label{eq:vacuum_potential}
E(\theta) = -\tilde{B}\, \sum_{i=1}^{N_F} M_i \, \cos\theta_i + \tilde{\mu}_{\Sigma}\cos\left(\sum_{k=1}^{N_F} \theta_k - \theta\right)\ .
\ee
In the previous equation we have used $\tilde{B} = B v \sqrt{2/N_F}$ and $\tilde{\mu}_\Sigma = 2 \mu_\Sigma (v/\sqrt{2 N_F})^{N_F}$. 
The Dashen equations give the minimum conditions
\be
\tilde{B} \, M_i\, \sin\theta_i =-\tilde{\mu}_\Sigma\, \sin\left(\sum_k \theta_k - \theta\right)\ , ~~~ i =1 , \dots , N_F\ .
\ee
We can find approximate solutions to the Dashen equations when the anomaly term dominates over the mass term, which is equivalent to assume that the mass of the $\eta'$ is larger than the masses of the mesons. In this limit we have
\be\label{eq:sum_vacuum_angles}
\sin\left(\sum_k \theta_k - \theta\right) \simeq 0 \ , ~~~ \Rightarrow ~~~ \theta \simeq \sum_k \theta_k\ .
\ee
The remaining Dashen equations can now be written as
\be\label{eq:Dashen_masses}
M_i \sin\theta_i = M_k \sin\theta_k \ , ~~~~ i \neq k \ .
\ee

In the case in which all the fermion masses are approximately of the same order the previous equation is  solved by $\theta_ i \simeq \theta_k \simeq \theta/N_F$. This means that in the limit considered, {\emph{i.e.}} dominance of the anomaly term over the mass term and approximate degeneracy of the fermion masses, the vacuum matrix is given by
\be
\Sigma_0 \simeq \frac{v\, e^{i \theta/N_F} }{\sqrt{2 N_F}} \mathbbm{1}\ .
\ee
If instead a hierarchy is present among the fermion masses the situation drastically changes. For definitiveness, let us consider the hierarchy $M_1 \ll M_i$, $i = 2 , \dots, N_F$ when the anomaly term dominates. Eq.~\eqref{eq:Dashen_masses} is now approximately solved by $\theta_1 \simeq \theta$ and $\theta_i \simeq 0$ for $i = 2 , \dots, N_F$. 
The conclusion is that, when a clear hierarchy is present among the fermion masses, only the lightest fermions contribute significantly to the vacuum energy. As last step, we remind that the inclusion of the relaxion field can be achieved simply promoting $\theta$ to a dynamical parameter,
\be
\theta \to \phi(x)/F\ .
\ee

We now move on with the discussion of the potential of the $\sigma$ particle, since it drives the phase transition discussed in Sec.~\ref{sec:GW_signal}. Having discussed the effect of the VEV of the light modes in the computation of the vacuum energy, see Eq.~\eqref{eq:vacuum_potential}, we now simply focus on the potential driven by $\sigma$. To write the full potential we implement thermal effects using the Truncated Full Dressing (TFD) procedure~\cite{curtin2016thermal}
\be
V_{\rm eff}(\sigma, T) = V(\sigma) + \sum_{i} \frac{T^{4}}{2\pi^{2}}\,n_{i} \,J_{B}\left(\frac{M^{2}_{i}(\sigma, T) }{T^{2}}\right)\ .
\ee
The first term, $V(\sigma)$, is the tree level potential as a function of the homogenous background field $\sigma$, and can be obtained from Eq.~\eqref{eq:strongly_interacting_generic}. Following refs.~\cite{Bai_2019,Croon:2019iuh} we consider the Coleman-Weinberg contribution already included in the tree level term, since its inclusion just renormalizes the tree-level couplings. Each thermal contribution depends on the multiplicity $n_i$, on the bosonic thermal integral
\be
J_B(x) = \int_0^\infty dy\, y^2 \log\left(1-e^{-\sqrt{x+ y^2}} \right)\ ,
\ee
and on the thermal masses $M_{i}(\sigma, T)$. In a sigma model with $N_F$ flavors they read
\al{
M^{2}_{\sigma}(\sigma, T) &= - m_{\Sigma}^{2}  + \frac{3}{2}\left(\lambda + \frac{\kappa}{N_{F}}\right) \sigma^{2}
- \mu_{\Sigma} (N_{F} -1) \left(\frac{\sigma}{\sqrt{2 N_{F}}}\right)^{N_{F}-2} +  \Pi(N_F)\\
M^{2}_{\eta'}(\sigma, T) &=- m_{\Sigma}^{2}  + \frac{1}{2}\left(\lambda + \frac{\kappa}{N_{F}}\right) \sigma^{2}
+ \mu_{\Sigma} (N_{F} -1) \left(\frac{\sigma}{\sqrt{2 N_{F}}}\right)^{N_{F}-2}+ 
\Pi(N_F) \\
M^{2}_{S^{a}}(\sigma, T) &=- m_{\Sigma}^{2}  + \frac{1}{2}\left(\lambda +  \frac{3\, \kappa}{N_{F}}\right) \sigma^{2}
+ \mu_{\Sigma} \left(\frac{\sigma}{\sqrt{2 N_{F}}}\right)^{N_{F}-2} + 
\Pi(N_F) \\
M^{2}_{\pi^{a}}(\sigma, T) &=- m_{\Sigma}^{2}  + \frac{1}{2}\left(\lambda +  \frac{ \kappa}{N_{F}}\right) \sigma^{2}
- \mu_{\Sigma} \left(\frac{\sigma}{\sqrt{2 N_{F}}}\right)^{N_{F}-2} + 
\Pi(N_F)\ ,
}
The terms proportional to $\mu_\Sigma$ are present only for $N_F \geq 3$. We have already included the ``Debye'' masses $\Pi(N_F)$ computed at one loop, the so called ``hard thermal loop"~\cite{curtin2016thermal}. 
 In a theory with $N_F$ flavor we obtain
\be
\Pi(N_F) =  \frac{T^{2}}{12}\bigg((N_{F}^{2} +1) \lambda + 2 N_{F}\, \kappa\bigg)\ .
\ee

\subsection{Explicit models}\label{app:explicit_models}
In the original paper~\cite{Graham2015} the back-reaction barrier is generated by the so-called $L+N$ model. This is a theory in which vector-like fermions charged under a new confining group are introduced. These fermions have EW quantum numbers to allow for Yukawa interactions with the Higgs boson. More specifically, the model consists in a vector-like pair $L$, $L^c$ (where $L$ has the same quantum numbers as the SM lepton doublet and $L^c$ has conjugated charges) and by a second vector-like pair $N$, $N^c$ of SM singlets. This is only one of the possibilities since, as we are now going to show, all models that allow for a Yukawa interaction with the Higgs doublet and in which there is a clear mass hierarchy can generate the required back-reaction. Before starting, let us remind the reader that in light of the Pisarski-Wilczek argument~\cite{Pisarski:1983ms} already mentioned in Sec.~\ref{sec:GW_signal}, we require the presence of at least 3 light flavors below the confinement scale to produce a strong first order phase transition. 

The most general Lagrangian we consider is
\be
{\cal L} = - \psi^c {\cal M}_\psi \psi - \chi^c {\cal M}_\chi \chi - H \psi^c {\cal Y} N\chi^c - H^\dag \chi^c {\cal Y'} \psi + h.c. 
\ee
where the quantum numbers of the vector-like fermions are such that it is possible to write the Yukawa interactions. Moreover, ${\cal M}_L$, ${\cal M}_N$, ${\cal Y}$ and ${\cal Y}'$ are matrices whose dimensions depend on the number of fermions. Suppose now there is a hierarchy $\mathcal{M}_\chi \ll \mathcal{M}_\psi$. Integrating out the heavy fermions we obtain the effective Lagrangian
\be
{\cal L}_{eff} = \chi^c \left( {\cal M}_\chi - {\cal Y}' \frac{1}{{\cal M}_\psi} {\cal Y}H^\dag H \right) \chi + h.c.
\ee
As explained in Sec.~\ref{app:general_setup}, the computation of the vacuum energy can be done analytically when all the fermions are approximately degenerate or when there is a clear mass hierarchy, in which case only the lightest fermions contribute. This allows us to conclude that the heavy $\psi$ states do not substantially contribute to the vacuum energy even if their masses happen to be below the confinement scale. The problem thus is to compute the eigenvalues of the $\chi$ mass matrix. To write approximate analytical formulas we  take all mass matrices to be proportional to the identity, $\mathcal{M}_\chi = m_\chi \mathbbm{1}$ and $\mathcal{M}_\psi = m_\psi \mathbbm{1}$ and all Yukawa matrices to be real with equal entries, $\mathcal{Y}^{(\prime)}_{ij} = y^{(\prime)} $. In this limit the vacuum energy is equal to
\be
E(\phi) \simeq - \tilde{B} \left| m_\chi - \frac{n_\psi n_\chi y y'}{m_\psi} H^\dag H \right| \cos\left(\frac{\phi}{F} \right)\ ,
\ee
where $n_\chi$ and $n_\psi$ are the number of $\chi$ and $\psi$ fermions. This equation justifies the form of the back-reaction used in Eq.~\eqref{eq:back-reaction}. Experimental limits on the $L+N$ model have been studied in~\cite{Beauchesne:2017ukw} for the situation in which only the $N$ flavor confines, and in~\cite{Antipin:2015jia,Barducci:2018yer} for the situation in which both $L$ and $N$ confine. Although relevant, none of these bounds put significant restrictions on the parameters of the linear sigma model used in the computation of the GW spectra.

Let us comment on two further points before concluding this section. There is a caveat to the above argument: when the fermions that form bound states carry EW quantum numbers, loop contributions to the effective Lagrangians can be important, see {\emph{e.g.}}~\cite{Antipin:2015jia}. We have assumed so far that such loops are negligible, but this is not necessarily the case. When loops are important the argument leading to Eq.~\eqref{eq:vacuum_Sigma_nf} fails, and the computation of the vacuum energy can in general be done only numerically. Finally, let us give some concrete example of models that can lead to strong first order phase transition. Focussing for simplicity only on variations of the original $L+N$ model used in Ref.~\cite{Graham2015}, we summarize the different possibilities in Fig.~\ref{fig:summary}. 
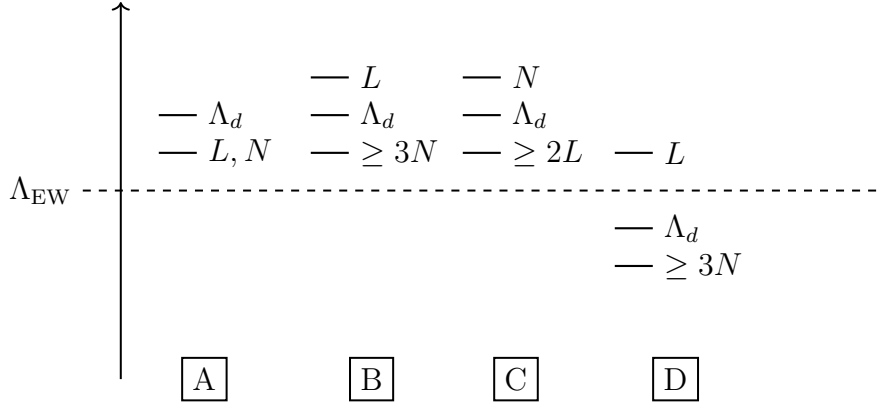
\begin{figure}[t!]
\begin{center}
\adjustbox{valign=m}{
 \begin{tikzpicture}[line width=0.75] 
\draw[->] (0,0) -- (0,5);
\draw[dashed] (-0.5,2.5) node [left]{$\Lambda_{\rm EW}$} -- (10, 2.5);
\draw[thick] (0.5, 3.5) -- (1., 3.5) node [right]{$\Lambda_d$};
\draw[thick] (0.5, 3) -- (1, 3) node [right] {$L, N$};
\draw[thick] (2.5, 3.5) -- (3, 3.5) node [right]{$\Lambda_d$};
\draw[thick] (2.5, 4) -- (3, 4) node [right]{$L$};
\draw[thick] (2.5, 3) -- (3, 3) node [right]{$\geq 3 N$};
\draw[thick] (4.5, 3.5) -- (5, 3.5) node [right]{$\Lambda_d$};
\draw[thick] (4.5, 3.) -- (5, 3.) node [right]{$\geq 2 L$};
\draw[thick] (4.5, 4) -- (5, 4) node [right]{$N$};
\draw[thick] (6.5, 3.) -- (7, 3.) node [right]{$L$};
\draw[thick] (6.5, 2.) -- (7, 2.) node [right]{$\Lambda_d$};
\draw[thick] (6.5, 1.5) -- (7, 1.5) node[right]{$\geq 3N$};
\node[draw] at (1.1,0) {A};
\node[draw] at (3.3,0) {B};
\node[draw] at (5.2,0) {C};
\node[draw] at (7.3,0) {D};
\end{tikzpicture}} 
\end{center}
\caption{\label{fig:summary_models} Summary of the models that can generate the back-reaction for the relaxion mechanism to work and, at the same time, satisfy the Pisarski-Wilczek condition. We show only variations of the $L+N$ model defined in Sec.~\ref{app:explicit_models}, although more general possibilities are possible.}
\end{figure}

\section{Successful relaxation of the EW scale with strongly interacting fermions}\label{sec:succ_relaxation}

We now describe in more  detail how the conditions listed in Sec.~\ref{sec:relaxation_inflation} are obtained. Some of these results have been presented in the literature in approximate form, but we give here more complete expressions. To keep the discussion generic, we will use the form of the back-reaction potential shown in Eq.~\eqref{eq:back-reaction}. Minimizing the potential in Eq.~\eqref{eq:potential} we obtain
\begin{align}
\label{eq:min1} \frac{\partial V}{\partial H} &= 0 \Rightarrow & v^2(\theta) & = -\frac{\Lambda^2 - \epsilon \Lambda F \, \theta -{\cal S} \, \mu_B^2 \cos\theta}{2 \lambda}\ , \\
\label{eq:min2} \frac{\partial V}{\partial \phi}  &= 0  \Rightarrow & r \epsilon \Lambda^3 F & = \left({\cal S} \mu_B^2 \sin\theta - \epsilon \Lambda F \right) v^2(\theta) - {\cal S}\Lambda_0^4 \,\sin\theta
\end{align}
where we have defined the dimensionless field $\theta = \phi/F$, $v^2(\theta)$ is the $\theta$-dependent Higgs minimum, and $\mathcal{S} = \mathrm{sign}[\mu_B^2 v^2(\theta) - \Lambda_0^4]$. As usual, the first equation applies when $v^2(\theta)$ is a positive quantity, otherwise the Higgs VEV vanishes. 
From the minimum equations we see that for small $\theta$ we have $v^2(\theta) = 0$, and $\mathcal{S}=-1$. From the equations of motion it follows that $\partial V /\partial\phi = 0$ corresponds to the stopping of the relaxion evolution. To avoid this, we need to make sure that the minimum equation has no solution while $v^2(\theta)=0$. This amounts to require
\be\label{eq:no_stopping}
r \epsilon \Lambda^3 F > \Lambda_0^4\ ,
\ee
which corresponds to Eq.~\eqref{eq:cond1}. As time passes and $\theta$ increases, the system reaches a critical value for the relaxion field in which $\epsilon \Lambda F \theta_c - \mu_B^2 \cos\theta_c = 2 \lambda \Lambda^2$ and EWSB is triggered. From Eq.~\eqref{eq:min1} we see that $v^2(\theta)$ starts growing essentially linearly with $\theta$. Looking at Eq.~\eqref{eq:min2}, and keeping into account Eq.~\eqref{eq:no_stopping}, we conclude that the right hand side must grow to guarantee the existence of a solution after EWSB. This can happen only if the factor multiplying $v^2(\theta)$ is positive. This requirement amounts to $\mathcal{S} = +1$ (at least around the EW scale), {\emph{i.e.}} $\mu_B^2 v^2_{EW} > \Lambda_0^4$, and to $\mu_B^2 > \epsilon \Lambda F$, which are Eq.~\eqref{eq:cond2}.

We now discuss the computation of the EW scale in terms of the parameters of the model. Once the EW minimum is reached for a value $\theta_0$ of the relaxion field we must have
\al{\label{eq:EW_min}
2 \lambda v^2_{EW} & = -\Lambda^2 + \epsilon \Lambda F \, \theta_0 +  \mu_B^2 \cos\theta_0\ , \\
 r \epsilon \Lambda^3 F & = \left( \mu_B^2 \sin\theta_0 - \epsilon \Lambda F \right) v^2_{EW} - \Lambda_0^4 \,\sin\theta_0\ .
}
Solving these equations for $\sin\theta_0$ and $\cos\theta_0$, and using $\sin^2\theta_0 + \cos\theta_0^2=1$ we can determine the value of $\theta_0$. The two solutions are
\al{\label{eq:theta_0}
\theta_0 &= \frac{\Lambda^2 + 2 \lambda v_{EW}^2}{\epsilon F \Lambda} \pm \frac{\mu_B^2}{\epsilon F \Lambda( \mu_B^2 v_{EW}^2 - \Lambda_0^4 )} \sqrt{\left( \mu_B^2 v_{EW}^2 - \Lambda_0^4 \right)^2 - (\epsilon F \Lambda^3)^2 \left( 1+ \frac{v_{EW}^2}{\Lambda^2} \right)}\ .
}
\begin{figure}[tb]
	\centering
	\includegraphics[scale=.48]{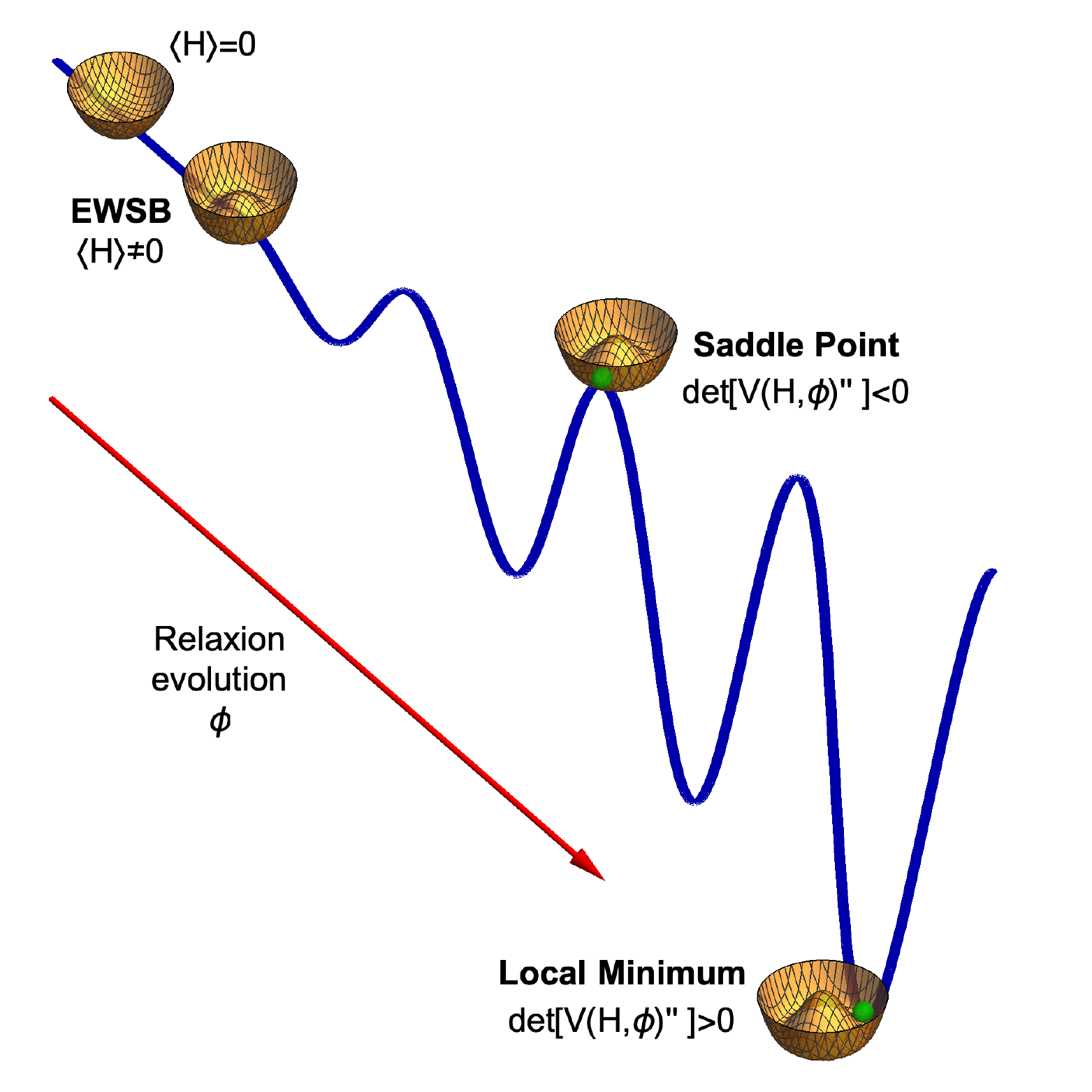}
	\caption{\label{fig:relaxion_higgs} Sketch of the electroweak relaxation mechanism. At the beginning of the relaxion evolution the Higgs squared mass parameter is positive and $\langle H \rangle = 0$. As $\phi$ evolves it drives $\langle H \rangle \neq 0$. Once the Higgs VEV is turned on, a back-reaction is generated, stopping the evolution. We also show the sign of the determinant of the matrix of second derivatives of the potential, important in deriving Eq.~\eqref{eq:DeltaV}. }
\end{figure}
The positive sign gives $\cos\theta_0 <0$, while the negative sign gives $\cos\theta_0 >0$. We now analyze the minimum conditions. Requiring $\det V'' >0$, where $V''$ is the matrix of second derivative of the potential, see the sketch in Fig.~\ref{fig:relaxion_higgs}, we obtain 
\be\label{eq:cos_min}
\cos\theta_0 > \frac{\epsilon^2 \Lambda^6 F^2 \left( \mu_B^2 + \frac{\Lambda_0^4}{\Lambda^2} \right)^2 }{2\lambda_H \left( \mu_B^2 v^2 - \Lambda_0^4 \right)^3}\ .
\ee
Since the right hand side is a positive quantity, we immediately conclude that the solution with $\cos\theta_0 <0$ corresponds to a saddle point, while the solution with $\cos\theta_0 >$ may be a local minimum in some region of parameter space. To determine this region we first translate Eq.~\eqref{eq:cos_min} in a maximum equation for $\sin\theta_0$. We then combine this maximum equation with Eq.~\eqref{eq:sin_0}, obtaining 
\be
\epsilon F \Lambda^3 < \left[ \mu_B^2 v_{EW}^2 - \Lambda_0^4\right] \left[ \left(1+\frac{v_{EW}^2}{\Lambda^2}\right)^2 + \frac{\mu_B^2 + \Lambda_0^4/\Lambda^2}{2\lambda_H (\mu_B^2 v_{EW}^2 - \Lambda_0^4)} \right]^{-1/2}\ .
\ee
Noticing that the term inside the square brackets is very close to one we end up with $\epsilon F \Lambda^3 \lesssim \mu_B^2 v_{EW}^2 - \Lambda_0^4$ as condition to guarantee the existence of a minimum. As for the saddle point solution, we notice that it corresponds to a minimum in the Higgs direction and a local maximum in the relaxion direction, see Fig.~\ref{fig:relaxion_higgs}. By focussing in the $\phi$ direction, the difference between the potential in the maximum and in the minimum reads
\al{\label{eq:DeltaV}
V_{max} - V_{min} & = \frac{2 (r \mu_B^2 \Lambda^2 + \Lambda_0^4)}{\mu_B^2 v_{EW}^2 - \Lambda_0^4} \sqrt{\left(\mu_B^2 v_{EW}^2 - \Lambda_0^4 \right)^2 - \epsilon^2 F^2 \Lambda^6 \left( r + \frac{v_{EW}^2}{\Lambda^2} \right)^2} \ .
}
Let us now go back to Eq.~\eqref{eq:EW_min}. The second equation can be used to compute
\be\label{eq:sin_0}
\sin \theta_0 = \left(r+\frac{v_{EW}^2}{\Lambda^2} \right) \frac{\epsilon F \Lambda^3 }{ \mu_B^2 v_{EW}^2 - \Lambda_0^4} \ .
\ee
Requiring $ |\sin \theta_0 | \leq 1$ we obtain 
\be
\Lambda^3  \left(r+\frac{v_{EW}^2}{\Lambda^2} \right) \leq  \frac{\mu_B^2 v_{EW}^2 - \Lambda_0^4}{\epsilon F}\ ,
\ee
which is the inequality of Eq.~\eqref{eq:upper_bound_Lambda}. \textit{In order to maximize the allowed value of $\Lambda$ we must be close to $\theta_0 = \pi/2 + 2 \pi \kappa$, with $\kappa$ an integer}. This agrees with the results of Ref.~\cite{Banerjee:2020kww}. Using this in Eq.~\eqref{eq:min1} allows us to compute how much the Higgs VEV changes as a function of $\kappa$. We obtain
\be
v^2(\kappa') - v^2(\kappa) = \frac{\pi \epsilon F \Lambda (\kappa' - \kappa)}{\lambda} \ .
\ee
In the case $\kappa' - \kappa = 1$, i.e. for two subsequent minima, we obtain
\be
\Delta v^2 \lesssim \frac{\pi}{\lambda} \frac{\mu_B^2 v_{EW}^2 - \Lambda_0^4}{\Lambda^2}\ ,
\ee
where we have used the condition on the parameter space required by the existence of a minimum. We then see that the change of the Higgs VEV between subsequent minima is very small.

\bibliographystyle{JHEP}
{\footnotesize
\bibliography{biblio_relaxion}}

\end{document}